\documentclass[fleqn,usenatbib]{mnras}

\usepackage{newtxtext,newtxmath}
\usepackage[T1]{fontenc}
\usepackage[bottom]{footmisc}
\DeclareRobustCommand{\VAN}[3]{#2}
\let\VANthebibliography\thebibliography
\def\thebibliography{\DeclareRobustCommand{\VAN}[3]{##3}\VANthebibliography}

\usepackage{graphicx}	% Including figure files
\usepackage{amsmath}
\usepackage{array,booktabs}
\usepackage{siunitx}
\title[Atmospheric Evolution of Small Planets]{Impact of M-dwarf Stellar Wind and Photoevaporation on the Atmospheric Evolution of Small Planets}

\author[Modi et al.]{
Ashini Modi,$^{1,2}$ \thanks{e-mail: ashinimodi@college.harvard.edu}
Raissa Estrela,$^{3}$
Adriana Valio$^{4}$
\\
$^{1}$Harvard University, Cambridge, MA 02138, USA\\
$^{2}$Caddo Parish Magnet High School, 1601 Viking Dr, Shreveport, LA 71101\\
$^{3}$Jet Propulsion Laboratory, California Institute of Technology, 4800 Oak Grove Drive, Pasadena, California 91109, USA\\
$^{4}$Center for Radio Astronomy and Astrophysics Mackenzie (CRAAM), Mackenzie Presbyterian University, Rua da Consolacao, 896, Sao Paulo, Brazil
}

\date{Accepted XXX. Received YYY; in original form ZZZ}

\pubyear{2015}

\DeclareUnicodeCharacter{2212}{-}

\begin{document}
\label{firstpage}
\pagerange{\pageref{firstpage}--\pageref{lastpage}}
\maketitle

\begin{abstract}
The evolution of a planet's atmosphere depends strongly on its host star's properties. When their host stars are younger, planets can experience stronger winds and EUV emissions. This is particularly true for planets orbiting M-dwarfs due to their close proximity to the host star. To determine if these planets retain an atmosphere, we consider the impacts from stellar wind and EUV fluxes in driving atmospheric escape throughout the planet's lifetime. For this, we determined the atmospheric mass loss due to stellar wind and photoevaporation on 4 planets in close orbit and 34 in their star's habitable zone (HZ). The M-dwarf host stars' wind velocity, density, and EUV flux were calculated through rotation period and X-ray flux scaling over time. The mass loss rate due to stellar wind and photoevaporation was then computed as a function of time and accumulated throughout the planet's age to determine the total atmospheric mass loss of the planet's initial H/He envelope. We find that for HZ planets at orbits $<$ 0.1 AU, stellar wind can only remove $\leq 1\%$ of the H/He envelope, while photoevaporation is essential for completely removing the H/He envelope of most targets. Moreover, due to either mechanism, most planets orbiting at $>$ 0.1 AU do not have their primordial envelope stripped. Overall, out of the 38 planets studied, 13 were predicted to have lost the primordial envelope due to photoevaporation, while 2 planets lost the envelope due to both stellar wind and photoevaporation.

\end{abstract}

\begin{keywords}
planets and satellites: atmospheres -- planetary systems -- stars: winds, outflows
\end{keywords}

%%%%%%%%%%%%%%%%%%%%%%%%%%%%%%%%%%%%%%%%%%%%%%%%%%

%%%%%%%%%%%%%%%%% BODY OF PAPER %%%%%%%%%%%%%%%%%%

\section{Introduction}

 Most of the recent attention in the exoplanet community has been focused on planets around M dwarfs, the most common type of star within the Milky Way. M dwarf stars are excellent targets for detecting small planets because the photometric transit depths are larger due to the smaller star-to-planet mass and size ratios, respectively \citep{shields2016habitability}. This imply that a transit signal of a small planet orbiting a M dwarf is relatively more significant than Sun-like stars. The Kepler mission has shown that these cool stars typically have terrestrial planets orbiting them, with the planetary radius ranging from 0.5-3R$_{\oplus}$. Also, most of the planets detected in the Habitable Zone (HZ) are orbiting M dwarfs \citep{shields2016habitability} in which HZ is defined as the circumstellar region in which a terrestrial planet can sustain liquid water on its surface \citep{Hart78,Kasting93,Koppa13}. 

However, being inside the HZ does not necessarily mean that the planets are conducive to life formation. Planets around M dwarfs are in much closer orbit and will experience more effects from the stellar winds and Extreme Ultraviolet Radiation (EUV or XUV) fluxes. This can significantly impact the retention of an atmosphere on the planet, which is one of the prerequisites for habitability \citep{lammer2009makes}. The star's evolution is therefore crucial in defining the outcome of the planetary atmosphere and, consequently, its capability to develop life \citep{davis2009evidence}.  

At the beginning of a planetesimal's lifetime, while a gas disk is still present, the rocky core can accumulate hydrogen and helium, forming a thick primordial atmosphere. More massive cores will accrete larger envelopes \citep{stokl2016dynamical}, resulting in pressures and temperatures incompatible with liquid water. However, this envelope can be removed by interaction with the planet's host star. Removing a primordial envelope can leave a bare core, requiring other processes such as volcanic or mantle out-gassing to establish a secondary atmosphere \citep{Kite20,swain2021detection}. If an envelope is not entirely removed but is reduced to a mass fraction of $<10^{-3}$, pressures potentially compatible with liquid water could exist \citep{pierrehumbert2011hydrogen}. Observational evidence of the host star's influence in shaping the planet's atmosphere can be seen in the planetary size distribution of small planets. This distribution exhibits an ``evaporation valley" \citep{fulton2017california}, which could be caused by the photoevaporation of the planetary atmosphere due to high stellar EUV fluxes in the first million years of the planetary lifetime, when the star is more active \citep{owen2013kepler,Jin14,lopez13,Howe14,Howe15,Rogers15,owen17,Van18}or by core-powered mass loss \citep{gupta2020signatures}.

Atmospheric erosion can also be induced by stellar wind, a stream of charged particles released from the host star's upper atmosphere (corona). In our Solar System, empirical evidence suggests that the solar wind constituted one of the dominant atmospheric loss mechanisms in early Earth, Mars, and Venus \citep{catling2017atmospheric}. Stellar winds can also vary in density, temperature, and speed over time. Assessing the possible atmospheric effects of the stellar wind over long timescales requires knowledge of the stellar activity evolution. At younger ages, stars typically lose more mass, creating intense stellar winds that could significantly impact the planet's conditions early on \citep{ribas2005evolution, griessmeier2009stellar}. However, as the star loses angular momentum over time, the rotation period will increase, and the X-ray flux will decay, decreasing the stellar mass loss \citep{irwin2007monitor}. 

Other studies such as \cite{cohen2015interaction}, \cite{tanaka2014atmospheric}, \cite{trammell2014magnetohydrodynamic}, and \cite{erkaev2017effect} have utilized 3D MHD models to compute the atmospheric loss of planets. However, because these more complex models are time-intensive, here we used a simple 1D analytical model to build a comprehensive population study of M-dwarf HZ planets. Previous works have used 1D models to study the effects of stellar winds \citep[e.g.][]{zendejas2010atmospheric} on planetary atmospheres through time. In this study, we use a similar approach by scaling the stellar rotation period with the X-ray flux. However, the evolution of planetary mass and radius are also considered. Further, by conducting a population analysis, we shed light on the transition from primordial to secondary atmospheres by analyzing whether close-in and HZ planets could have lost a primordial envelope during their lifetime. %However, this simpler model reached comparable results with MHD simulations.

Given the importance of the stellar wind and photoevaporation in defining the possible outcome of the planetary atmosphere,  we analyze the influence of these mechanisms in driving atmospheric mass loss of the initial atmospheric content on small planets around M-dwarf stars. Here we assume time-dependent and un-magnetized planetary escape models due to stellar wind and photoevaporation, reflecting the system's evolution with age. The stellar wind and photoevaporation atmospheric escape models are described in Section 2. Our results from the modeling are illustrated in Section 3. In Section 4, we discuss the model and results further. In Section 5, we present the conclusions.\\

\section{Methods}

\subsection{Planet and stellar population in study}
Since the presence of an atmosphere is essential for habitability, in this work, we have analyzed 34 planets in the HZ of their M dwarf host star. Additionally, we included 4 planets (GJ 1132 b, GJ 1214 b, Kepler-138d, and K2-3c) that are in close-in orbits. From the combined sample, 8 of the planets currently have observations with the Hubble Space Telescope (HST) using the transmission spectroscopy technique (GJ-1132b, LHS 1140b, K2-18b, GJ 1214 b, and Trappist-1 (d-g)).
%All planets in our sample are orbiting M dwarf stars.}
Understanding the processes that shape these planets' atmosphere evolution is crucial to contextualize the current observations. Based on their density, mass, and radius, these planets can be classified as terrestrial, sub-Neptunes, or in the transition region between terrestrials and sub-Neptunes. The parameters of the planets and their host stars are shown in Tables~\ref{tab:planet} and~\ref{tab:star}, respectively. All observational parameters (mass, radius, orbital distance) of each planet were taken from the ``Planetary systems composite'' data of the NASA Exoplanet Archive. It is important to note that the data drawn from the NASA exoplanet archive planetary systems composite data may not be self-consistent if the parameters were drawn from multiple sources. The data in this archive combines parameters obtained from different sources, and when there is no mass or radius available, it calculates these values with a mass-radius relationship. We indicate in Table~\ref{tab:planet} the targets in our sample that had their parameters (mass or radius) estimated using a mass-radius relationship.

\begin{table}
\resizebox{\columnwidth}{!}{%
\begin{tabular}{ccccc}
\hline
\hline
Planet name   & \begin{tabular}[c]{@{}c@{}}Planet mass\\ ($\rm M_\oplus$)\end{tabular} & \begin{tabular}[c]{@{}c@{}}Planet radius\\ ($\rm R_\oplus$)\end{tabular} & \begin{tabular}[c]{@{}c@{}}Semi-major\\ (AU)\end{tabular} & \begin{tabular}[c]{@{}c@{}}Equilibrium\\ Temperature\\ (K)\end{tabular} \\ \hline
GJ 1061 c     & 1.74 $\pm$ 0.23                                                              & 1.18*                                                                 & 0.035$\pm$0.001                                                       & 304$\pm$10.10                                                            \\
GJ 1061 d     & 1.64$\pm$ 0.23                                                                & 1.16*                                                                 & 0.054$\pm$0.001                                                       & 245$\pm$8.13                                                            \\
GJ 1132 b      & 1.66$\pm$ 0.23                                                               & 1.30$\pm$0.05                                                                  & 0.0153$\pm$0.0005                                                      & 529$\pm$9                                                            \\
GJ 1214 b      & 8.17$\pm$0.43                                                              & 2.74$\pm$0.05                                                                 & 0.01490$\pm$0.00026                                                     & 596$\pm$19                                                            \\
GJ 163 c       & 6.80$\pm$0.9                                                               & 2.50*                                                                 & 0.1254$\pm$0.0001                                                      & 304$\pm$8.7                                                            \\
GJ 180 c       & 6.40$\pm$3.7                                                               & 2.41*                                                                 & 0.129$\pm$0.017                                                      & 289$\pm$8.5                                                            \\
GJ 229 A c    & 7.27$\pm$0.67                                                               & 2.60*                                                                 & 0.3842$\pm$0.0051                                                      & 244$\pm$7.9                                                            \\
GJ 273 b      & 2.89$\pm$0.27                                                               & 1.51*                                                                 & 0.91101$\pm$0.000019                                                      & 292$\pm$4.2                                                            \\
GJ 3293 d     & 7.60$\pm$1.05                                                               & 2.67*                                                                 & 0.19394$\pm$0.000017                                                      & 251$\pm$3.4                                                            \\
GJ 357 d       & 6.10$\pm$1.0                                                               & 2.34*                                                                 & 0.204$\pm$0.015                                                      & 226$\pm$6                                                            \\
GJ 667 C c    & 3.80$\pm$1.5                                                               & 1.77*                                                                 & 0.125$\pm$0.012                                                      & 278$\pm$3.9                                                            \\
GJ 667 C e    & 2.70$\pm$1.6                                                               & 1.45*                                                                 & 0.213$\pm$0.019                                                      & 213$\pm$3                                                            \\
GJ 667 C f    & 2.70$\pm$1.4                                                               & 1.45*                                                                 & 0.156$\pm$0.017                                                      & 249$\pm$3.5                                                            \\
GJ 682 b      & 4.40$\pm$3.7                                                               & 1.93*                                                                 & 0.080$\pm$0.014                                                      & 215$\pm$4.66                                                            \\
GJ 832 c       & 5.40$\pm$1.0                                                               & 2.18*                                                                 & 0.161$\pm$0.017                                                      & 286$\pm$7.92                                                            \\
K2-18 b        & 8.92$\pm$1.7                                                               & 2.37$\pm$0.22                                                                 & 0.1429$\pm$0.0065                                                      & 303$\pm$15                                                            \\
K2-288 B b     & 4.27*                                                               & 1.90$\pm$0.3                                                                 & 0.164$\pm$0.030                                                      & 234$\pm$22                                                            \\
K2-3 c         & 3.10$\pm$1.08                                                               & 1.77$\pm$0.10                                                                 & 0.1345$\pm$0.0016                                                      & 374$\pm$7   \\                                               
K2-3 d         & 2.80*                                                               & 1.53$\pm$0.11                                                                 & 0.2097$\pm$0.0070                                                      & 282$\pm$24                                                            \\
K2-72 e        & 2.21*                                                               & 1.29$\pm$0.14                                                                 & 0.106$\pm$0.013                                                      & 307$\pm$7.4                                                            \\
K2-9 b         & 5.69*                                                               & 2.25$\pm$0.96                                                                 & 0.091$\pm$0.016                                                      & 316$\pm$67                                                            \\
Kepler-1229 b & 2.54*                                                               & 1.40$\pm$0.13                                                                 & 0.3006$\pm$0.0091                                                      & 217$\pm$19                                                            \\
Kepler-186 f  & 1.71*                                                               & 1.17$\pm$0.08                                                                 & 0.432$\pm$0.171                                                      & 198$\pm$4.76                                                               \\
Kepler-138 d   & 0.64$\pm$0.67                                                               & 1.21$\pm$0.08                                                                 & 0.12781$\pm$0.00456                                                      & 343$\pm$4.38                                                            \\
Kepler-1649 c  & 1.20*                                                               & 1.06$\pm$0.15                                                                 & 0.0649*                                                      & 303$\pm$20                                                            \\
Kepler-1652 b  & 3.19*                                                               & 1.60$\pm$0.18                                                                 & 0.1654$\pm$0.0042                                                      & 276$\pm$27                                                            \\
Kepler-296 e   & 2.96*                                                               & 1.53$\pm$0.27                                                                 & 0.169$\pm$0.029                                                      & 282$\pm$18                                                            \\
Kepler-296 f   & 3.89*                                                               & 1.80$\pm$0.31                                                                 & 0.255$\pm$0.043                                                      & 229$\pm$15                                                            \\
Kepler-705 b   & 5.10*                                                               & 2.11$\pm$0.10                                                                 & 0.2320$\pm$0.0037                                                      & 269$\pm$21                                                            \\
LHS 1140 b     & 6.38$\pm$0.46                                                               & 1.64$\pm$0.05                                                                 & 0.0957$\pm$0.0019                                                      & 222$\pm$4                                                            \\
Proxima Cen b & 1.27$\pm$0.19                                                               & 1.08*                                                                 & 0.0485$\pm$0.0051                                                      & 257$\pm$6                                                            \\
Ross 128 b    & 1.40$\pm$0.21                                                               & 1.11*                                                                 & 0.0496$\pm$0.0017                                                      & 309$\pm$5.81                                                            \\
TOI-700 d      & 1.57*                                                               & 1.14$\pm$0.06                                                                 & 0.1633$\pm$0.0026                                                      & 278$\pm$8                                                            \\
TRAPPIST-1 d  & 0.39$\pm$0.01                                                               & 0.79$\pm$0.01                                                                 & 0.02227$\pm$0.00019                                                      & 296$\pm$6                                                            \\
TRAPPIST-1 e  & 0.69$\pm$0.02                                                               & 0.92$\pm$0.01                                                                 & 0.02925$\pm$0.00250                                                      & 258$\pm$5                                                            \\
TRAPPIST-1 f  & 1.04$\pm$0.03                                                               & 1.05$\pm$0.01                                                                 & 0.03849$\pm$0.00033                                                      & 225$\pm$4                                                            \\
TRAPPIST-1 g  & 1.32$\pm$0.04                                                               & 1.13$\pm$0.02                                                                 & 0.04683$\pm$0.00040                                                      & 204$\pm$4                                                            \\
Wolf 1061 c    & 3.41$\pm$0.43                                                               & 1.66*                                                                 & 0.0890$\pm$0.0031                                                      & 306$\pm$4.4                                                            \\ \hline
\end{tabular}%
}
\caption{Planetary properties (planet mass, radius, and semi-major axis) of our sample. This data was obtained from the ``Planetary systems composite data” tab on the NASA exoplanet archive, which reports parameters from different sources, selecting the most precise value from the literature or, if the most precise value can not be uniquely identified, the most recent value. All values with * next were estimated using a mass-radius relationship by the NASA exoplanet archive. Equilibrium temperatures were calculated from \protect\cite{mendez2017equilibrium} assuming a bond albedo of 0. }
\label{tab:planet}
\end{table}

\begin{table}
\begin{tabular}{cccc}
\hline
\hline
Host star name & Spectral type & \begin{tabular}[c]{@{}c@{}}Stellar radius\\ ($\rm R_\odot$)\end{tabular} & \begin{tabular}[c]{@{}c@{}}Stellar mass\\ ($\rm M_\odot$)\end{tabular} \\ \hline
GJ 1061        & M5.5V         & 0.16                                                                 & 0.12                                                               \\
GJ 1132        & M3.5V         & 0.21                                                                 & 0.18                                                               \\
GJ 1214        & M4.5V         & 0.22                                                                 & 0.15                                                               \\
GJ 163         & M5.5V         & 0.41                                                                 & 0.4                                                                \\
GJ 180         & M2 V          & 0.41                                                                 & 0.43                                                               \\
GJ 229 A       & M1.5V         & 0.46                                                                 & 0.58                                                               \\
GJ 273         & M3.5V         & 0.29                                                                 & 0.29                                                               \\
GJ 3293        & M2.5V         & 0.4                                                                  & 0.42                                                               \\
GJ 357         & M2.5V         & 0.34                                                                 & 0.34                                                               \\
GJ 667 C       & M1.5V         & 0.42                                                                 & 0.33                                                               \\
GJ 682         & M3.5V         & 0.3                                                                  & 0.27                                                               \\
GJ 832         & M1.5V         & 0.44                                                                 & 0.45                                                               \\
K2-18          & M2.5V         & 0.41                                                                 & 0.36                                                               \\
K2-288 B       & M3V           & 0.32                                                                 & 0.33                                                               \\
K2-3           & M0V           & 0.56                                                                 & 0.6                                                                \\
K2-72          & M2.5V         & 0.33                                                                 & 0.27                                                               \\
K2-9           & M2.5V         & 0.31                                                                 & 0.3                                                                \\
Kepler-1229    & M0V           & 0.51                                                                 & 0.54                                                               \\
Kepler-186     & M1V           & 0.52                                                                 & 0.54                                                               \\
Kepler-138     & M1V           & 0.44                                                                 & 0.52                                                               \\
Kepler-1649    & M5V           & 0.23                                                                 & 0.2                                                                \\
Kepler-1652    & M2V           & 0.38                                                                 & 0.4                                                                \\
Kepler-296     & M2V           & 0.48                                                                 & 0.5                                                                \\
Kepler-705     & M2V           & 0.51                                                                 & 0.53                                                               \\
LHS 1140       & M4.5V         & 0.21                                                                 & 0.19                                                               \\
Proxima Cen    & M5.5V         & 0.14                                                                 & 0.12                                                               \\
Ross 128       & M4V           & 0.2                                                                  & 0.17                                                               \\
TOI-700        & M2V           & 0.42                                                                 & 0.41                                                               \\
TRAPPIST-1     & M8V           & 0.12                                                                 & 0.09                                                               \\
Wolf 1061      & M3.5V         & 0.31                                                                 & 0.29                                                               \\ \hline
\end{tabular}
\caption{Host star properties used for model calculations}
\label{tab:star}
\end{table}

\subsection{Time-dependent atmospheric escape}
We have carried out the time-dependent atmospheric erosion considering two escape processes: stellar wind and photoevaporation. These models are detailed in the subsections below. For both processes, we compute the cumulative atmospheric mass loss backward in time (Section~\ref{sec:evol}), together with the evolution of the planetary mass and radius. To determine if the planet could have lost its envelope, we determine a threshold which is given by an estimated primordial envelope (Section~\ref{sec:threshold}). 

\subsubsection{Stellar wind velocity}
The stellar wind velocity was determined from the extrapolation of the ``Parker wind''  applied to the cool dwarfs \citep{parker1958dynamics}. The hydrodynamic equation of a wind assuming a steady flow is given by:
\begin{equation}
   \rm \frac{dP}{dr} + \frac{GM_{s} \rho}{r^{2}} = 0
\end{equation}

\noindent For a steadily expanding wind, the momentum equation that describes the acceleration of the gas due to a pressure gradient and gravity:

\begin{equation}
   \rm v \frac{dv}{dr} + \frac{1}{\rho} \frac{dP}{dr} + \frac{GM_{s}}{r^{2}} = 0
\end{equation}

\noindent where $v$ is velocity, $r$ is orbital radius, $G$ is the gravitational constant, $M$ is mass, $\rho$ is density, and P is pressure. We are assuming a perfect gas, P = R$\rm \rho$T/$\rm \mu$ where R is the gas constant and $\mu$ is the mean atomic weight. 

\begin{figure*}
 \includegraphics[width=0.32\textwidth]{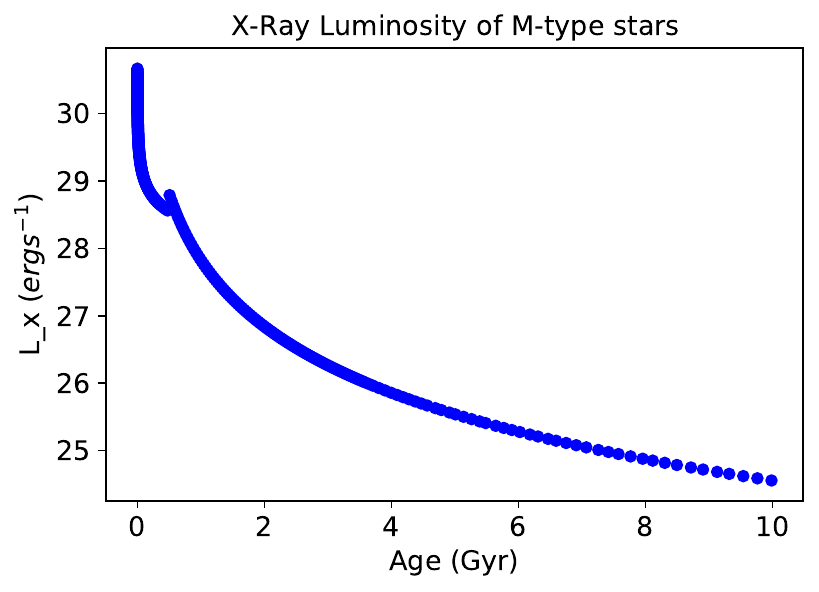}
 \includegraphics[width=0.33\textwidth]{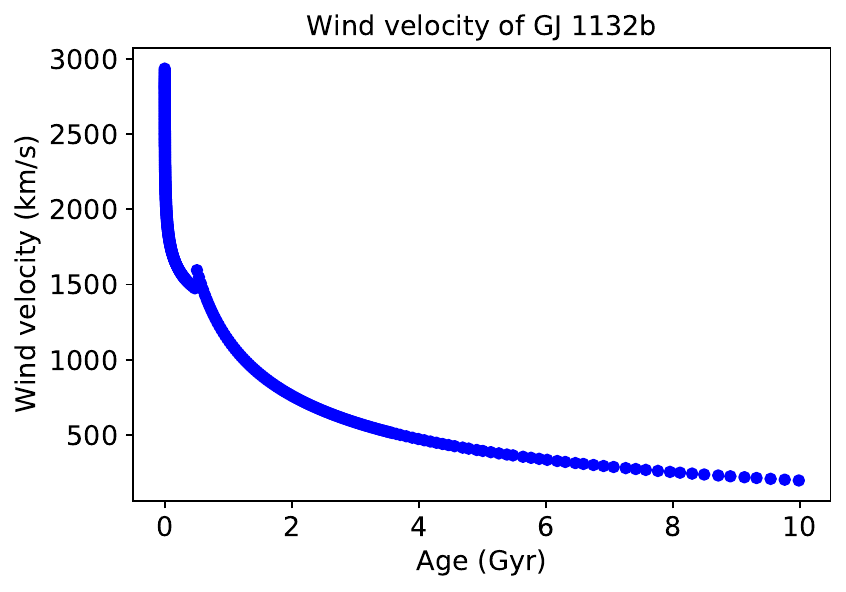}
 \includegraphics[width=0.32\textwidth]{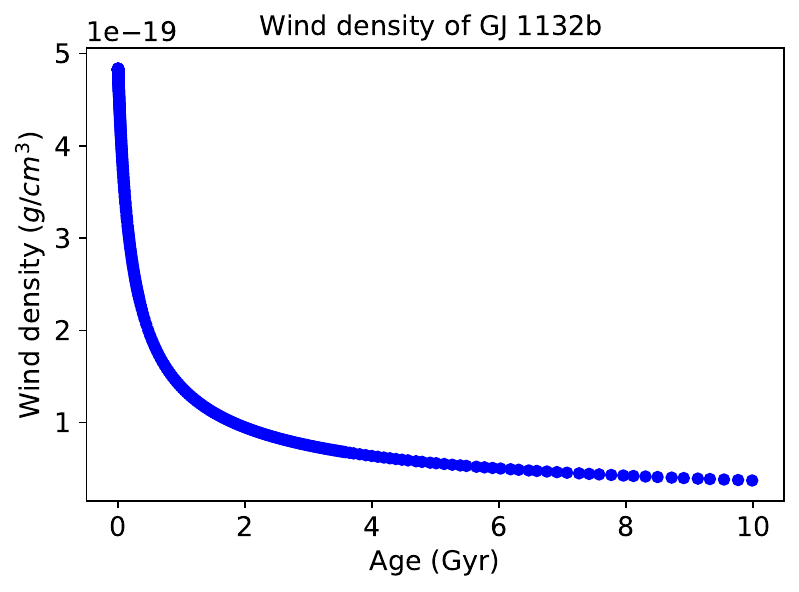}
      %\hspace*{-1.5cm}
   \caption{Estimated X-ray luminosity (left) based on observations of M dwarf stars from {\protect\cite{engle2018rotation}}, wind velocity (middle), and wind density (right) as a function of age for GJ 1132 b. The discontinuity in the wind velocity is due to the two different L$_{\rm x}$ relationships for $<$0.5 Gyr and $>$0.5 Gyr.}
    \label{fig:lx}
\end{figure*}

There are several solutions for the momentum equation, however, the only physical one is the sub-sonic, which passes through the critical point r=r$_{\rm c}$ at the sonic radius $\rm r_{c} = GM/2v_{c}^{2}$, and increases for larger distances. 
The terminal velocity of the wind (v$_{\rm w}$) is equal to the constant sound speed $\rm v_{c} = \sqrt{RT/\mu}$, in which $T$ is the coronal temperature. In this solution, the wind is spherically symmetric and isothermal and thus has a nearly constant temperature through the inner regions of the corona. To compute the coronal temperature of the stars, we use a relationship between the coronal temperature and the X-ray flux (F$_{\rm x}$) of the star given by T = 0.11 F$_{x}^{0.26}$ \citep{johnstone2015coronal}, where $T$ is in MK and $F_{\rm x}$ in erg s$^{-1}$ cm$^{−2}$. The stellar X-ray flux (F$_{\rm x}$) is calculated through time using the X-rays luminosity (L$_{\rm x}$) from \cite{engle2018rotation} based on observations of M dwarfs stars, as shown in Figure~\ref{fig:lx}. The relationship between F$_{\rm x}$ and  L$_{\rm x}$ is given by $\rm F_{x} = L_{x}/4\pi R_\star^2$.

The following relationships based on data from \cite{engle2018rotation} were used to scale X-ray luminosity with time, where $\tau$ is the age of the star:
\begin{align}
 \rm  L_x=-0.308\ \log(\tau)+28.333 \rm \qquad (for \; ages \; \textless 0.5 Gyr) \\
  \rm L_x=-1.4214\ \log(\tau)+27.826 \rm \qquad (for \; ages \; \textgreater 0.5 Gyr)
\end{align}

By rearranging and integrating the momentum equation results in:
\begin{equation}
   \rm \left(\frac{v}{v_{c}}\right)^{2} - \ln\left(\frac{v}{v_{c}}\right)^{2} = 4\ \ln\left(\frac{r}{r_{c}}\right) + 4\frac{r_{c}}{r} + C
\end{equation}

\noindent where $C$ is the constant of the integration and is equal to -3 in the transonic solution. For large $\rm r$ ($\rm r >> r_{c}$), we have $\rm v_{w} \approx 2v_c [\ln(\frac{r}{r_{c}})]^{\frac{1}{2}}$, whereas for small $\rm r$ ($\rm r << r_{c}$), $\rm v_{w} \approx (v_{c}{\rm e}^{\frac{3}{2}})^{\frac{-2r_{c}}{r}}$. An example of scaling the X-ray fluxes and wind velocity overtime for GJ 1132 b is shown in Fig.~\ref{fig:lx}.

\subsubsection{Stellar wind density}
Wind densities of stars other than our Sun have not been observationally constrained yet. As an approximation, the electron density derived from observations of the Sun by \cite{Leblanc98} is used to model the solar wind throughout interplanetary space. The density in terms of the radial distance (r) from the Sun is given by:
\begin{equation}
 \rm   n(r) = 3.3\times 10^5\ r^{2} + 4.1\times 10^6\ r^{-4} + 8.0\times 10^7\ r^{-6} ({\rm cm}^{-3})
\end{equation}

This equation was adapted so that data near Earth and estimates of the coronal density, $N_{s}$, matched better:
\begin{equation}
  \rm  n(r) = 7\times 10^4\ r^{-1.67} + 4.1\times 10^6\ r^{-4} + N_{s}\ r^{-6} ({\rm cm}^{-3})
    \label{eq:dens}
\end{equation}
\noindent where $N_{s} = 2.6\times 10^7$ cm$^{-3}$.

To estimate the wind density over time for other stars,  Eq.~\ref{eq:dens} was scaled with the rotation period of the star, as a proxy for  the stellar age \citep{Vidotto20}:
\begin{equation}
   \rm \rho(r,t) = n(r) m_{\rm H} \left(\frac{Prot_{\odot}}{Prot_{star}} \right)^{0.6}
\end{equation}

Stars lose their angular momentum with time, resulting in the lengthening of their rotation periods with age. Therefore, to provide a more accurate analysis of atmospheric escape through time, it is crucial to consider the evolution of the rotation period. We use the rotation period-age relationship from \cite{engle2018rotation} according to each M dwarf type, where $\tau$ is the age of the star: 
\begin{align}
    \rm P_{rot} = (2\pm 3) + (15.8 \pm 0.6)\ \tau \rm \qquad (M2.5-M6 \; dwarfs)
    \label{eq:Prot}
\end{align}
\begin{align}
    \rm P_{rot} = (-5 \pm 6) + (17\pm  5) \ \tau^{0.669} \rm \qquad (M0-M1 \; dwarfs)
    \label{eq:Prot2}
\end{align}

These relationships are less precise for younger ages ($<$ 0.8 Gyrs) because there is a wide scatter in the range of rotation period rates for early ages, which prevents the establishment of a clear trend in the rotation period-age plane \citep{popinchalk2021evaluating}. This leads to unrealistic values of rotation period at ages $<$ 0.2 Gyrs using Eq.~\ref{eq:Prot2}, and therefore we perform our calculations for M0V stars starting at 0.2 Gyr. Stars that do not fall into the spectral ranges in Eqs.~\ref{eq:Prot} \&~\ref{eq:Prot2} were grouped into the closest category. For stars with spectral type $>$ M6V, we kept the rotation period constant, as they likely do not experience the same decay. These stars appear to show rapid rotations through intermediate and advanced ages \citep{mohanty2003rotation, reiners2010volume}. For example, Trappist-1 (an active M8-type star) has a rotation period of 1.40$\pm$0.05 days \citep{gillon2016temperate} and an age of 7.6 ±2.2 Gyr \citep{burgasser2017age}.  An example of scaling wind density over time is shown in Fig.~\ref{fig:lx} for the star GJ 1132. The stellar mass loss over time due to the stripping of the stellar envelope by the winds are also considered. To model this effect, we compute the mass loss of the star at the current age and add the mass back in time to the stellar mass at intervals of $\approx$ $\rm 1\times 10^{7}$ Gyr.

\subsubsection{Atmospheric mass loss due to stellar wind}

Once the wind density and velocity have been estimated, the rate of the planetary atmospheric mass loss due to the stellar wind is calculated by \citep{zendejas2010atmospheric}:
\begin{equation}
   \dot{m} = \rm 2 \pi R_p^{2} \alpha \rho_{w} v_{w}
    \label{eq:massloss}
\end{equation}
where $\alpha$ is an entrainment coefficient with a value ranging from 0.01 to 0.3, R$_{\rm p}$ is the planet radius, v$_{\rm w}$, and $\rho_w$ are the wind velocity (km s$^{-1}$) and density (g cm$^{-3}$), respectively. The highest value of $\alpha=0.3$ was adopted to determine maximum mass loss. This same value of entrainment, $\sim$0.3, was estimated for Venus by \cite{bauer2013planetary}.

It is worth noting that this model assumes an isothermal stellar wind; however, stellar winds are not, in fact, perfectly isothermal. Other studies have used more complex models that consider the magnetohydrodynamic of the Alfvén waves to derive a detailed structure of the wind energetics \citep{Mesquita2020global}. Their MHD numerical simulation of stellar winds yields that the winds reach isothermal temperatures very quickly. Comparison made by the authors with the Parker wind models resulted in good agreement for the high-beta regime.

\subsubsection{Atmospheric mass loss due to photoevaporation}

To better evaluate atmospheric erosion in our sample, we compare the results of the stellar wind model to the energy-limited photoevaporation model. Photoevaporation probably played a crucial role in shaping the atmospheric evolution of terrestrial planets in the first 100 Myrs of their lifetime \citep{owen2013kepler, lopez2013role, howe2014mass, howe2015evolutionary, rogers2015most, owen2017evaporation, van2018asteroseismic}. In a H/He atmosphere, high energy radiation from the host star can heat the atmospheric gas, affecting the atmospheric stability. This is particularly true for close-in planets such as those orbiting M-dwarf stars. XUV radiation below 91.2 nm can be directly absorbed by atomic hydrogen because this is the limit that the photon has enough energy ($h_0 \geq 13.6 eV$) to ionize an H atom. The mass loss rate due to XUV radiation can be estimated using the approach from \cite{kubyshkina2020coupling}:

\begin{equation}
 \dot{m} = \frac{\epsilon \pi F_{EUV} R_p R_{XUV} ^2}{G M_p K}
    \label{eq:massloss1}
\end{equation}
where $\epsilon$ is the heating efficiency which was taken to be 0.1, a value shown to be appropriate for terrestrial and sub-Neptune planets \citep{owen2012planetary,owen2017evaporation,lopez2012thermal,chen2016evolutionary}, $\rm F_{EUV}$ is the XUV flux received by the planet, and $\rm R_{EUV}$ is the effective absorption radius for XUV photons. The term $\rm K$ accounts for the fact that atmospheric material only escapes out to the Roche-lobe height and represents the fractional gravitational potential energy difference between the planetary surface and the Roche-lobe height. The factor $\rm K$ is calculated using the  equation from \cite{erkaev2007roche}:
\begin{equation}
\rm K=1-\frac{3}{2\xi}+\frac{1}{2\xi^3} \text{  with  } \xi=(\frac{M_p}{3M_s})^{1/3}\frac{a}{R_p}
\label{eq:K}
\end{equation}
where $\rm M_p$ is the planet mass, $\rm M_s$ is the stellar mass, $\rm a$ is the semi-major axis, and $\rm R_{p}$ is the planet radius. The effective absorption radius for XUV photons, R$_{XUV}$, was calculated using the formalism from \cite{kubyshkina2020coupling}. 
\begin{equation}
\rm R_{XUV}=R_{p}+H \ln\left( \frac{P_{photo}}{P_{XUV}} \right)
\label{eq:effrad}
\end{equation}
\noindent where $\rm H$ is the atmospheric scale height, $\rm P_{photo}$ is the surface pressure, and $\rm P_{XUV}$ is the pressure at the XUV absorption level. The atmospheric scale height is given by H = $\rm K_{b} T_{eq}/\mu g$ where $\rm K_{b}$ is the Boltzmann constant, $\rm \mu$ is the mean molecular weight of an atmosphere in solar composition ($\rm \mu$ = 2.3 amu), $g$ is the surface gravity, and $\rm T_{eq}$ is the calculated equilibrium temperature. The pressure, $\rm P_{XUV}$ in $\rm kg m^{-1} s^{-2}$, was estimated using the following equation:
\begin{equation}
\rm P_{XUV}=\frac{m_HGM_p}{\sigma_{v0}R^2_p}
\label{eq:effp}
\end{equation}
where G is the gravitational constant in units of $\rm m^3 kg^{-1} s^{-2}$, m$_{H}$ is the mass of hydrogen in kg, respectively, and $\rm \sigma_{v0}$ is the absorption cross-section. $\rm M_p$ is in units of kg and $\rm R_p$ is in units of m. The effective radius, the radius of the planet where XUV flux is absorbed, is calculated by scaling the planetary radius ($\rm R_p$) with time, as described in Section~\ref{sec:evol}.
%$F_{\rm EUV}$ was calculated using the relationship between L$_{x}$ and F$_{\rm x}$. 
F$_{\rm EUV}$ was calculated using the power law relation with F$_{\rm x}$,  from \cite{chadney2015xuv}:
\begin{equation}
    \rm \log F_{EUV}= 2.63+0.58\log F_x.
\end{equation}

\subsubsection{Evolution of planetary mass and radius}\label{sec:evol}
\begin{figure}
 \centering
 \includegraphics[width=0.4\textwidth]{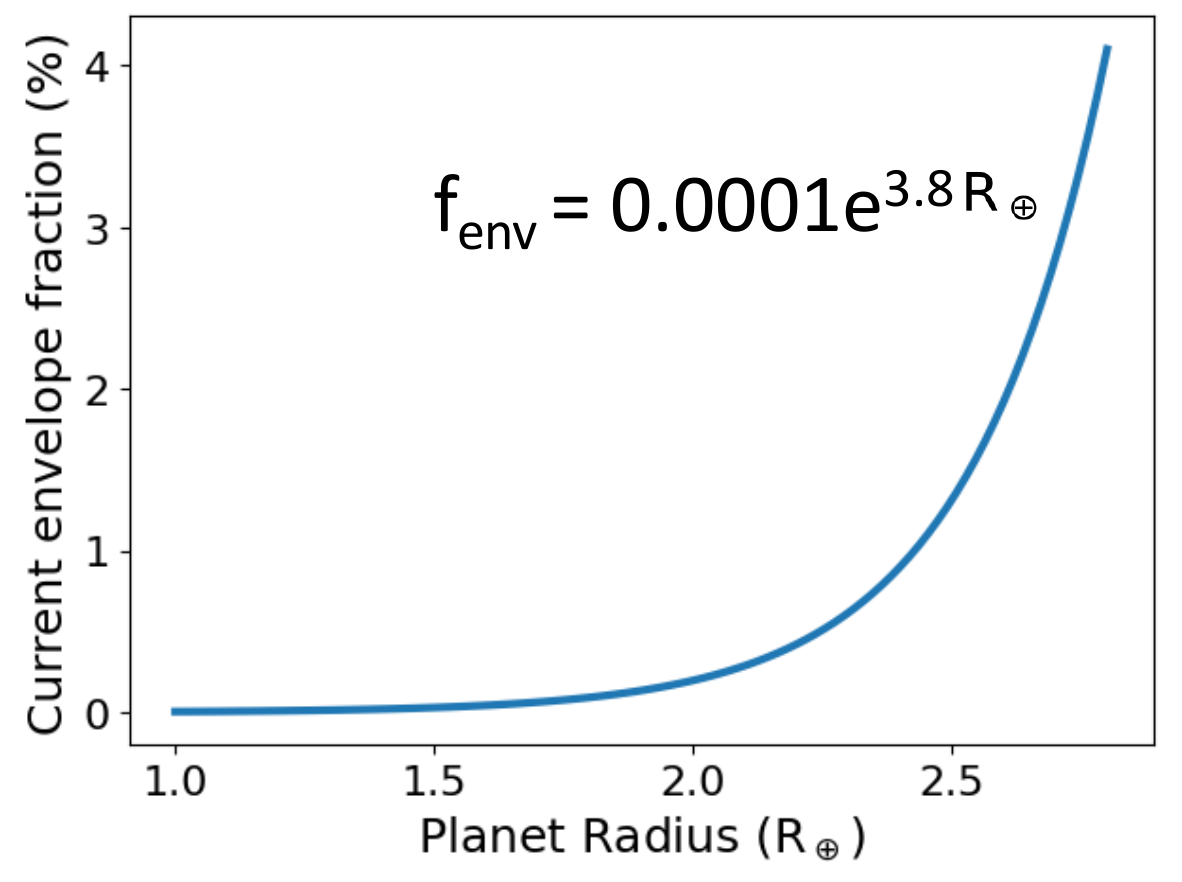}
   \caption{Relationship between current envelope fraction and planetary radius derived by exponentially fitting an observational dataset presented in \protect\cite{lopez2014understanding}. The empirically derived fit equation $\rm f_{env}= 0.0001e^{3.8R_{\oplus}}$ is shown for planets with radius up to 3 R$_{\oplus}$. The determined current envelope fraction serves as a fundamental starting point for our subsequent estimations of atmospheric mass loss.}
   \label{fig:envfrac}
 \end{figure}

To account for the changes in planetary mass and radius caused by the stripping of the H/He envelope, we consider both photoevaporation and stellar wind loss mechanisms backward in time by following these steps:

\begin{enumerate}
\item First, we determine the initial envelope fraction, f$_{\rm env}$, of the planet, by using the current envelope fraction as our starting point since our calculations are backward in time. This value is estimated by performing an exponential fit to the relationship between the envelope fraction (\%) and planet radius, using the observational sample from Figure 8 of \cite{lopez2014understanding}. Our fit is presented in Figure \ref{fig:envfrac}. It is worth noting that even a relatively small amount of H/He has a significant impact on the planetary radius \citep{lopez2014understanding}.

\item Next, we calculate the mass loss of the planet at its current age using Equation~\ref{eq:massloss} for stellar wind or Equation~\ref{eq:massloss1} for photoevaporation. We then add this estimated mass loss to the current planet mass (M$_{\rm p}$) yielding the mass of the planet at a previous age (M$_{\rm bf}$):

\begin{equation}
\rm   M_{bf} = M_p + \dot{m}
\end{equation}

\item We then re-estimate the radius of the planet at the previous ages and recalculate the mass loss. The radius of the planet at the previous ages (R$_{\rm bf}$) is obtained by first estimating the envelope radius (R$\rm _{env}$) associated with its envelope fraction which is given by the mass added back (estimated envelope lost) and then adding it to the core radius: 

\begin{equation}
    \rm R_{bf} = R_{core} + R_{env}
\end{equation}

For this estimation, we use the R$\rm_{env}$  relation from \cite{lopez2014understanding} (where $\rm R_{env} = R_{p} - R_{core} - R_{atm}$) for an enhanced opacity and assuming that the planets have cores with Earth-like composition, which is given by:

\begin{equation}
   \rm R_{env}= 2.06\ R_\oplus \left( \frac{M_p}{M_\oplus}\right)^{-0.21} \left(\frac{f_{env}} {5\%}\right)^{0.59} \left(\frac{F_p}{F_\oplus} \right)^{0.044}\left (\frac{\tau}{5Gyr}\right)^{-0.18}
\label{eq:radius}
\end{equation}

\noindent where $\rm f_{\rm env}$ is the planet envelope fraction and $\rm F_p$ is the insolation the planet receives, and $\rm M_p$ is the planet mass. These parameters are all scaled with time. The insolation, $\rm F_p$, is estimated using the relation from \cite{weiss2014mass}:

\begin{equation}
    \rm \frac{F_p}{F_\oplus} = \left( \frac{R_\star}{R_\odot} \right)^2 \left( \frac{T_{eff}}{5778} \right)^4 \left( \frac{1}{(1-e^2)} \right)^{0.5}
\end{equation}
\noindent where $\rm T_{eff}$ and $\rm \frac{R_\star}{R_\odot}$ evolve with time using evolutionary models from \cite{baraffe2015new}.

\item Finally, the total mass lost from the planet's atmosphere was computed by integrating the mass loss accumulated over time.
\end{enumerate}

\subsubsection{Primordial envelope threshold}\label{sec:threshold}
Once the total atmospheric mass lost due to stellar wind and photoevaporation has been computed,
to determine if the planet has lost its envelope, we define a threshold set by an estimated primordial envelope mass. Based on our assumptions, the planet will have lost its envelope if the accumulated mass lost in 5 Gyr is $\geq$ than the primordial envelope mass threshold. We estimate the primordial envelope by using a relationship from \cite{Lee19} that gives the amount of gas mass accreted given the core mass of the planet:
\begin{equation}
   \rm \frac{M_{gas}}{M_{core}}=0.09 \left(\frac{\sum_{neb}}{13gcm^{-2}} \right)^{0.12} \left(\frac{M_{core}}{20M_\oplus} \right)^{0.17} \left(\frac{t}{0.1Myr} \right)^{0.4}
    \end{equation}
\noindent where $\rm \sum_{neb}$ is the nebular mass surface density and $t$ is the timescale to accrete the gas. In the case of super-Earths and mini-Neptunes, they emerge during the late stage when the nebula is gas-poor, and the timescale to accrete gas would take approximately 5.7 Myrs \citep{Lee19}. We adopted a value of 13 g cm$^{-2}$ for $\rm \sum_{neb}$. $\rm M_{core}$ was estimated by subtracting the mass of the current envelope from the planet current mass, where the current envelope is estimated using the relationship from \cite{lopez2014understanding} as shown in Fig.~\ref{fig:envfrac}. The estimated primordial envelopes are indicated in Table~\ref{tab:env}.

\begin{table}
\begin{tabular}{ccc}
\hline
\hline
Planet name   & Current envelope (\%) & Primordial envelope (\%) \\ \hline
GJ 1061 c     & 0.01*                  & 0.51                     \\
GJ 1061 d     & 0.01*                  & 0.46                     \\
GJ 1132 b      & $0.01\pm1\times10^{-4}$                  & 0.47                     \\
GJ 1214 b      & $4.96\pm1\times10^{-4}$                  & 5.49                     \\
GJ 163 c       & 1.31*                  & 6.99                     \\
GJ 180 c       & 0.93*                  & 6.37                     \\
GJ 229 A c    & 1.92*                  & 7.70                     \\
GJ 273 b      & 0.03*                  & 1.69                     \\
GJ 3293 d     & 2.51*                  & 8.17                     \\
GJ 357 d       & 0.72*                  & 5.91                     \\
GJ 667 C c    & 0.08*                  & 2.69                     \\
GJ 667 C e    & 0.02*                  & 1.51                     \\
GJ 667 C f    & 0.02*                  & 1.51                     \\
GJ 682 b      & 0.15*                  & 3.44                     \\
GJ 832 c       & 0.39*                  & 4.85                     \\
K2-18 b        & $0.80\pm2.3\times10^{-4}$                  & 11.25                    \\
K2-288 B b     & $0.13\pm3.1\times10^{-4}$                  & 3.27                     \\
K2-3 c         & 0.08*                  & 1.90                     \\
K2-3 d         & $0.03\pm5.5\times10^{-4}$                  & 1.60                     \\
K2-72 e        & $0.01\pm1.52\times10^{-4}$                  & 1.07                     \\
K2-9 b         & $0.51\pm1.70\times10^{-4}$                  & 5.28                     \\
Kepler-1229 b & $0.02\pm7.47\times10^{-4}$                  & 1.36                     \\
Kepler-186 f  & $0.01\pm1.64\times10^{-4}$                  & 0.50                     \\
Kepler-138 d   & $0.01\pm1.35\times10^{-4}$                  & 0.09                     \\
Kepler-1649 c  & $0.01\pm1.77\times10^{-4}$                  & 0.27                     \\
Kepler-1652 b  & $0.04\pm1.98\times10^{-4}$                  & 2.00                     \\
Kepler-296 e   & $0.03\pm2.79\times10^{-4}$                  & 1.76                     \\
Kepler-296 f   & $0.09\pm3.24\times10^{-4}$                  & 2.80                     \\
Kepler-705 b   & $0.30\pm1.46\times10^{-4}$                  & 4.41                     \\
LHS 1140 b     & $0.05\pm1.19\times10^{-4}$                  & 6.49                     \\
Proxima Cen b & 0.01*                  & 0.30                     \\
Ross 128 b    & 0.01*                  & 0.35                     \\
TOI-700 d      & $0.01\pm1.27\times10^{-4}$                  & 0.43                     \\
TRAPPIST-1 d  & $0.01\pm1.04\times10^{-4}$                  & 0.04                     \\
TRAPPIST-1 e  & $0.01\pm1.05\times10^{-4}$                  & 0.11                     \\
TRAPPIST-1 f  & $0.01\pm1.05\times10^{-4}$                  & 0.21                     \\
TRAPPIST-1 g  & $0.01\pm1.06\times10^{-4}$                  & 0.32                     \\
Wolf 1061 c    & 0.05*                  & 2.24                     \\ \hline
\end{tabular}

\caption{Primordial envelope fractions relative to planet total mass estimated using the relationship from \protect\cite{Lee19} and current envelope fractions estimated using the relationship between $\%$ H/He and planet radius for water-rich planets from \protect\cite{lopez2014understanding}. Values with an * next to them indicate that mass/radius values were obtained with a mass-radius relationship in the NASA exoplanet archive.}
\label{tab:env}
\end{table}

\section{Results}
We estimated the evolution of the density and velocity of the stellar wind for 30 M-type stars hosting planets. For the M0-1V stars, at 5 Gyr, we find an average of 400 km s$^{-1}$ and $1.5\times 10^{-22}$ g cm$^{-3}$ for the wind velocity and density. While for stars of spectral type $>$M2V, these values at 5 Gyr are 490 km s$^{-1}$ an $1.2\times10^{-20}$ g cm$^{-3}$, respectively. The measured wind velocity and density of the solar wind near Earth orbit today are around 450 km s$^{-1}$ and $1\times 10^{-23}$ g cm$^{-3}$. 

Once the stellar properties were estimated, the planetary atmospheric mass loss rate due to stellar wind was computed as a function of time. Moreover, we also computed the atmospheric mass loss rate due to photoevaporation. Since there is a significant increase in stellar activity at younger ages, the mass loss rate for both mechanisms significantly drops off after 0.1 Gyr. This decrease at later ages indicates that photoevaporation and stellar wind loss mechanisms are more efficient in the first 100 Myr of the stellar lifetime. The results for the evolution of the stellar properties at time-points of $\approx 0.1/0.2$ Gyr and 5 Gyr are shown in Table~\ref{tab:s}.

\begin{center}
\begin{table*}
\centering
%\hspace*{-2cm}
\footnotesize
 \begin{tabular}{cccccc|ccccc}
\hline
\hline
age           & \multicolumn{5}{c}{0.1 Gyr (M2-8 V)  or 0.2 Gyr (M0-M1.5 V)}                                  & \multicolumn{5}{c}{5 Gyr}                                           \\ \hline
parameter     & $\rm v_w$           & $\rm \rho_w$           & $\rm T_{cor}$  & $\rm Prot$   & $\rm F_x$                  & $\rm v_w$           & $\rm \rho_w$           & $\rm T_{cor}$ & $\rm Prot$   & $\rm F_x$                  \\ \hline
unit          & (km s$^{-1}$) & (g cm$^{-3}$) & (MK)  & (days) & (erg s$^{-1}$ cm$^{-3}$) & (km s$^{-1}$) & (g cm$^{-3}$) & (MK) & (days) & (erg s$^{-1}$ cm$^{-2}$) \\ \hline
GJ 1061 c     & 2444.91      & $1.47\times 10^{-20}$     & 12.22 & 3.83   & $7.38\times 10^{7}$              & 628.06       & $2.35\times 10^{-21}$     & 1.50 & 81.49  & $2.32\times 10^{4}$              \\
GJ 1061 d     & 1973.87      & $7.99\times 10^{-21}$     & 12.22 & 3.83   & $7.38\times 10^{7}$              & 643.75       & $1.28\times 10^{-21}$     & 1.50 & 81.49  & $2.32\times 10^{4}$              \\
GJ 1132 b      & 1735.85      & $3.51\times 10^{-19}$     & 10.45 & 3.83   & $4.05\times 10^{07}$              & 394.29       & $5.83\times 10^{-20}$     & 1.28 & 81.49  & $1.26\times 10^{4}$              \\
GJ 1214 b      & 1733.53      & $6.23\times 10^{-19}$     & 10.32 & 3.83   & $3.85\times 10^{7}$              & 408.32       & $8.43\times 10^{-20}$     & 1.27 & 81.49  & $1.21\times 10^{4}$              \\
GJ 163 c       & 1461.33      & $1.28\times 10^{-21}$     & 7.40  & 3.83   & $1.07\times 10^{7}$              & 372.35       & $2.05\times 10^{-22}$     & 0.91 & 81.49  & $3.36\times 10^{3}$              \\
GJ 180 c       & 918.30       & $1.18\times 10^{-21}$     & 7.40  & 3.83   & $1.07\times 10^{7}$             & 369.65       & $1.89\times 10^{-22}$     & 0.91 & 81.49  & $3.36\times 10^{3}$              \\
GJ 229 A c    & 1506.62      & $9.50\times 10^{-20}$     & 6.97  & 0.71   & $8.51\times 10^{6}$              & 402.95       & $3.92\times 10^{-23}$     & 0.86 & 43.64  & $2.67\times 10^{3}$              \\
GJ 273 b      & 1627.67      & $2.42\times 10^{-21}$     & 8.81  & 3.83   & $2.10\times 10^{7} $             & 422.14       & $3.86\times 10^{-22}$     & 1.08 & 81.49  & $6.58\times 10^{3}$              \\
GJ 3293 d     & 1532.09      & $5.19\times 10^{-22}$     & 7.46  & 3.83   & $1.10\times 10^{7} $             & 404.66       & $8.28\times 10^{-23}$     & 0.92 & 81.49  & $3.46\times 10^{3} $             \\
GJ 357 d       & 1662.90      & $4.69\times 10^{-22}$     & 8.19  & 3.83   & $1.59\times 10^{7}$              & 450.28       & $7.48\times 10^{-23}$     & 1.01 & 81.49  & $4.97\times 10^{3}$              \\
GJ 667 C c    & 1478.18      & $3.36\times 10^{-21}$     & 7.27  & 0.71   & $1.00\times 10^{7}$              & 418.41       & $2.92\times 10^{-22}$     & 1.01 & 43.64  & $5.12\times 10^{3} $             \\
GJ 667 C e    & 1560.72      & $1.15\times 10^{-21}$     & 7.27  & 0.71   & $1.00\times 10^{7}$              & 458.28       & $9.96\times 10^{-23}$     & 1.01 & 43.64  & $5.12\times 10^{3}$              \\
GJ 667 C f    & 1512.96      & $2.15\times 10^{-21}$     & 7.27  & 0.71   & $1.00\times 10^{7}$              & 435.39       & $1.87\times 10^{-22}$     & 1.01 & 43.64  & $5.12\times 10^{3}$              \\
GJ 682 b      & 1742.66      & $6.32\times 10^{-22}$     & 8.74  & 3.83   & $2.03\times 10^{7} $             & 475.20       & $1.01\times 10^{-22}$     & 1.07 & 81.49  & $6.28\times 10^{3} $             \\
GJ 832 c       & 1317.47      & $2.66\times 10^{-21}$     & 6.26  & 0.71   & $5.62\times 10^{6}$              & 362.49       & $2.36\times 10^{-22}$     & 0.88 & 43.64  & $2.92\times 10^{3}$              \\
K2-18 b        & 1498.74      & $9.60\times 10^{-22}$     & 7.38  & 3.83   & $1.06\times 10^{7} $             & 391.28       & $1.54\times 10^{-22}$     & 0.91 & 81.49  & $3.35\times 10^{3}$              \\
K2-288 B b     & 1660.77      & $7.29\times 10^{-22}$     & 8.42  & 3.83   & $1.76\times 10^{7}$              & 443.88       & $1.16\times 10^{-22}$     & 1.03 & 81.49  & $5.45\times 10^{3}$              \\
K2-3 c         & 1314.41      & $8.68\times 10^{-22}$     & 6.29  & 3.83   & $3.88\times 10^{6}$              & 372.48       & $2.28\times 10^{-22}$     & 0.77 & 43.64  & $1.78\times 10^{3}$              \\
K2-3 d         & 1267.51      & $3.94\times 10^{-22}$     & 6.29  & 3.83   & $3.88\times 10^{6}$              & 339.02       & $1.03\times 10^{-22}$     & 0.77 & 43.64  & $1.78\times 10^{3}$              \\
K2-72 e        & 1558.21      & $3.09\times 10^{-21}$     & 8.28  & 3.83   & $1.65\times 10^{7}$              & 398.87       & $4.91\times 10^{-22}$     & 1.01 & 81.49  & $5.09\times 10^{3}$              \\
K2-9 b         & 1591.59      & $2.48\times 10^{-21}$     & 8.56  & 3.83   & $1.88\times 10^{7}$              & 408.87       & $3.94\times 10^{-22}$     & 1.05 & 81.49  & $5.81\times 10^{3}$              \\
Kepler-1229 b & 1337.05      & $1.87\times 10^{-22}$     & 5.80  & 0.71   & $4.20\times 10^{6} $             & 381.99       & $3.69\times 10^{-23}$     & 0.81 & 43.64  & $2.15\times 10^{3}$              \\
Kepler-186 f  & 1380.29      & $2.77\times 10^{-22}$     & 5.72  & 0.71   & $3.99\times 10^{6}$             & 404.52       & $2.41\times 10^{-23}$     & 0.80 & 43.64  & $2.04\times 10^{3}$              \\
Kepler-138 d   & 1313.27      & $1.07\times 10^{-21}$     & 7.10  & 0.71   & $9.16\times 10^{6}$              & 342.79       & $2.84\times 10^{-22}$     & 0.87 & 43.64  & $2.90\times 10^{3}$              \\
Kepler-1649 c  & 1771.28      & $4.93\times 10^{-21}$     & 9.96  & 3.83   & $3.36\times 10^{7}$              & 550.81       & $7.85\times 10^{-22}$     & 1.22 & 81.49  & $1.04\times 10^{4}$              \\
Kepler-1652 b  & 1540.51      & $7.16\times 10^{-22}$     & 7.68  & 3.83   & $1.24\times 10^{7} $             & 402.85       & $1.14\times 10^{-22}$     & 0.94 & 81.49  & $3.82\times 10^{3}$              \\
Kepler-296 e   & 1404.28      & $6.86\times 10^{-22}$     & 6.82  & 3.83   & $7.84\times 10^{6} $             & 357.15       & $1.09\times 10^{-22}$     & 0.83 & 81.49  & $2.42\times 10^{3}$              \\
Kepler-296 f   & 1467.60      & $3.00\times 10^{-22}$     & 6.82  & 3.83   & $7.84\times 10^{6}$              & 387.03       & $4.77\times 10^{-23}$     & 0.83 & 81.49  & $2.42\times 10^{3} $             \\
Kepler-705 b   & 1414.37      & $3.61\times 10^{-22}$     & 6.59  & 3.83   & $6.88\times 10^{6}$              & 368.70       & $5.78\times 10^{-23}$     & 0.81 & 81.49  & $2.17\times 10^{3}$              \\
LHS 1140 b     & 1890.78      & $2.28\times 10^{-21}$     & 10.38 & 3.83   & $3.94\times 10^{7} $             & 516.09       & $3.65\times 10^{-22}$     & 1.28 & 81.49  & $1.25\times 10^{4}$              \\
Proxima Cen b & 2021.67      & $1.47\times 10^{-20}$     & 12.88 & 3.83   & $9.03\times 10^{7}$              & 648.33       & $2.35\times 10^{-21}$     & 1.58 & 81.49  & $2.84\times 10^{4}$              \\
Ross 128 b    & 1846.15      & $8.89\times 10^{-21}$     & 10.83 & 3.83   & $4.64\times 10^{7}$              & 574.28       & $1.42\times 10^{-21}$     & 1.33 & 81.49  & $1.46\times 10^{4}$              \\
TOI-700 d      & 1482.71      & $7.36\times 10^{-22}$     & 7.26  & 3.83   & $9.98\times 10^{6}$              & 385.80       & $1.18\times 10^{-22}$     & 0.89 & 81.49  & $3.14\times 10^{3}$              \\
TRAPPIST-1 d  & 2107.91      & $1.18\times 10^{-18}$     & 14.07 & 1.40   & $1.27\times 10^{8}$              & 662.06       & $1.13\times 10^{-19}$     & 1.72 & 1.40   & $3.94\times 10^{4} $             \\
TRAPPIST-1 e  & 2113.93      & $6.06\times 10^{-19}$    & 14.07 & 1.40   & $1.27\times 10^{8}$              & 677.67       & $5.80\times 10^{-20}$    & 1.72 & 1.40   & $3.94\times 10^{4}$              \\
TRAPPIST-1 f  & 2214.40      & $2.85\times 10^{-19}$     & 14.07 & 1.40   & $1.27\times 10^{8}$              & 604.87       & $2.73\times 10^{-20}$     & 1.72 & 1.40   & $3.94\times 10^{4}$              \\
TRAPPIST-1 g  & 2254.34      & $1.78\times 10^{-19}$     & 14.07 & 1.40   & $1.27\times 10^{8}$              & 622.67       & $1.70\times 10^{-20}$     & 1.72 & 1.40   & $3.94\times 10^{4} $             \\
Wolf 1061c    & 1588.33      & $2.84\times 10^{-21}$     & 8.60  & 3.83   & $1.91\times 10^{7}$              & 406.81       & $4.53\times 10^{-22}$     & 1.05 & 81.49  & $5.94\times 10^{3}$              \\ \hline
\end{tabular}
\caption{Estimated properties of their host stars and the stellar wind at 0.1 Gyr (M2-M8V), 0.2 Gyr (M0-M1.5V), and at 5 Gyr. The value of $F_x$ was taken at planetary orbit.}
\label{tab:s}
\end{table*}
\end{center}
\vspace*{-7mm}

The results showing the comparison of the evolution of planet properties (mass, radius, total mass lost) at $\sim$0.1 Gyr and 5 Gyr due to stellar wind and photoevaporation are shown in Tables~\ref{tab:p2} and~\ref{tab:p3}, respectively.
Furthermore, the mass loss of the atmospheres leveled off at $\approx$ 2 Gyr in the system lifetime. This suggests that if an atmosphere of a planet has not been stripped before 2 Gyr, it is unlikely that it will be in the future. Also, it means that any later secondary outgassed atmosphere will not be caused by stellar wind stripping because of the decreased stellar activity.

\begin{center}
\begin{table*}
\resizebox{\textwidth}{!}{%
 \begin{tabular}{cccccc|ccccc}
\hline
\hline
age           & \multicolumn{5}{c|}{0.1 or 0.2 Gyr}                                                                                 & \multicolumn{5}{c}{5 Gyr}                                                                                           \\ \hline
parameter     & $\dot{m}$     & $\rm M_p$         & $\rm R_p$         & mass lost  & \begin{tabular}[c]{@{}c@{}}envelope \\ fraction lost\end{tabular} & $\dot{m}$     & $\rm M_p$         & $R_p$         & mass lost  & \begin{tabular}[c]{@{}c@{}}envelope \\ fraction lost\end{tabular} \\ \hline
unit          & (g/s)    & (M$_{\oplus}$) & (R$_{\oplus}$) & (M$_{\oplus}$) & (\%)                                                              & (g/s)    & (M$_{\oplus}$) & (R$_{\oplus}$) & (M$_{\oplus}$) & (\%)                                                              \\ \hline

GJ 1061 c     & $3.39\times 10^{6}$     & 1.74        & 1.24        & $1.85\times 10^{-5}$        & $2.08\times 10^{-1}$                                                          & $1.58\times 10^{5}$     & 1.74        & 1.18        & $1.98\times 10^{-5}$       & $2.23\times 10^{-1}$                                                          \\
GJ 1061 d     & $1.78\times 10^{6}$     & 1.64        & 1.22        & $9.82\times 10^{-6}$        & $1.30\times 10^{-1}$                                                          & $8.50\times 10^{4}$     & 1.64        & 1.16        & $1.05\times 10^{-5}$        & $1.39\times 10^{-1} $                                                         \\                                                    
GJ 1132 b      & $7.65\times 10^{7}$     & 1.66        & 1.28        & $3.36\times 10^{-4}$        & $4.30\times 10^{0}$                                                          & $2.18\times 10^{6}$     & 1.66        & 1.14        & $3.70\times 10^{-4}$        & $4.74\times 10^{0}$                                                          \\
GJ 1214 b      & $1.22\times 10^{9}$     & 8.14        & 4.19        & $3.74\times 10^{-3}$        & $8.37\times 10^{-1}$                                                          & $1.78\times 10^{7}$     & 8.14        & 2.60        & $4.56\times 10^{-3}$        & $1.02\times 10^{0}$                                                         \\
GJ 163 c       & $1.44\times 10^{6}$     & 6.80        & 3.17        & $5.70\times 10^{-6}$        & $1.20\times 10^{-3}$                                                          & $3.70\times 10^{4}$     & 6.80        & 2.52        & $6.48\times 10^{-6}$        & $1.36\times 10^{-3} $                                                         \\
GJ 180 c       & $1.14\times 10^{6}$     & 6.40        & 2.95        & $4.66\times 10^{-6}$        & $1.14\times 10^{-3}$                                                          & $3.10\times 10^{4}$     & 6.40        & 2.41        & $5.26\times 10^{-6}$        & $1.29\times 10^{-3} $                                                         \\
GJ 229 A c    & $9.46\times 10^{7}$     & 7.27        & 2.95        & $6.78\times 10^{-5}$        & $1.21\times 10^{-2}$                                                          & $5.63\times 10^{3}$     & 7.27        & 2.17        & $1.10\times 10^{-4}$        & $1.97\times 10^{-2} $                                                         \\
GJ 273 b      & $7.66\times 10^{5}$     & 2.89        & 1.60        & $3.74\times 10^{-6}$        & $7.65\times 10^{-3}$                                                          & $2.82\times 10^{4}$     & 2.89        & 1.51        & $4.05\times 10^{-6}$        & $8.30\times 10^{-3} $                                                        \\
GJ 3293 d     & $7.71\times 10^{5}$     & 7.60        & 3.57        & $2.90\times 10^{-6}$        & $4.67\times 10^{-4}$                                                          & $1.82\times 10^{4}$     & 7.60        & 2.67        & $3.35\times 10^{-6}$        & $5.39\times 10^{-4} $                                                        \\
GJ 357 d       & $4.62\times 10^{5}$     & 6.10        & 2.79        & $1.99\times 10^{6}$        & $5.52\times 10^{-4}$                                                          & $1.41\times 10^{4}$     & 6.10        & 2.34        & $2.22\times 10^{-6}$        & $6.17\times 10^{-4} $                                                                   \\
GJ 667 C c    & $3.18\times 10^{8}$     & 3.80        & 1.92        & $2.52\times 10^{-4}$        & $2.46\times 10^{-1}$                                                          & $2.93\times 10^{4}$    & 3.80        & 1.77        & $3.61\times 10^{-4}$        & $3.54\times 10^{-1}$                                                         \\
GJ 667 C e    & $7.17\times 10^{7}$     & 2.70        & 1.52        & $5.69\times 10^{-5}$        & $1.40\times 10^{-1}$                                                          & $7.34\times 10^{3}$     & 2.70        & 1.45        & $8.02\times 10^{-5}$        & $1.97\times 10^{-1} $                                                        \\
GJ 667 C f    &  $1.31\times 10^{8}$     & 2.70        & 1.52        & $1.04\times 10^{-4}$        & $2.55\times 10^{-1}$                                                          & $1.31\times 10^{4}$     & 2.70        & 1.45        & $1.47\times 10^{-4} $       & $3.61\times 10^{-1} $                                                        \\
GJ 682 b      & $2.13\times 10^{8}$     & 4.40        & 2.12        & $1.68\times 10^{-4}$        & $1.11\times 10^{-1}$                                                          & $1.98\times 10^{4}$     & 4.40        & 1.93        & $2.43\times 10^{-4}$        & $1.60\times 10^{-1} $                                                       \\
GJ 832 c       & $4.06\times 10^{8}$     & 5.40        & 2.56        & $3.18\times 10^{-4}$        & $1.21\times 10^{-1}$                                                          & $3.20\times 10^{4}$     & 5.40        & 2.21        & $4.71\times 10^{-4}$        & $1.80\times 10^{-1} $                                                        \\
K2-18 b        & $8.76\times 10^{5}$     & 8.92        & 2.82        & $3.47\times 10^{-6}$        & $3.45\times 10^{-4}$                                                          & $2.57\times 10^{4}$     & 8.92        & 2.37        & $3.74\times 10^{-6} $       & $3.73\times 10^{-4} $                                                     \\
K2-288 B b     & $3.99\times 10^{5}$     & 4.27        & 2.08        & $1.89\times 10^{-6}$        & $1.35\times 10^{-3}$                                                          & $1.42\times 10^{4}$     & 4.27        & 1.90        & $2.06\times 10^{-6}$        & $1.48\times 10^{-3} $                                                        \\
K2-3 c         & $2.17\times 10^{8}$     & 3.10        & 1.93        & $1.58\times 10^{-4}$        & $2.69\times 10^{-1}$                                                          & $2.03\times 10^{4}$     & 3.10        & 1.77        & $2.32\times 10^{-4}$        & $3.94\times 10^{-1} $                                                        \\
K2-3 d         & $7.75\times 10^{7}$     & 3.10        & 1.74        & $5.66\times 10^{0}$        & $1.14\times 10^{-1}$                                                          & $7.30\times 10^{3}$     & 3.10        & 1.65        & $8.20\times 10^{-5}$        & $1.66\times 10^{-1}      $                                                 \\
K2-72 e        & $6.58\times 10^{5}$     & 2.21        & 1.34        & $3.27\times 10^{-6}$        & $1.38\times 10^{-2}$                                                          & $2.49\times 10^{4}$     & 2.21        & 1.29        & 3.54$\times 10^{-6}$        & $1.50\times 10^{-2}  $                                                        \\
K2-9 b         & $2.06\times 10^{6}$     & 5.69        & 2.83        & $8.09\times 10^{-6}$        & $2.69\times 10^{-3}$                                                          & $5.19\times 10^{4}$     & 5.69        & 2.25        & $9.21\times 10^{-6}$        & $3.07\times 10^{-3}$                                                          \\
Kepler-1229 b & $5.46\times 10^{4}$     & 2.54       & 1.46        & $2.72\times 10^{-7}$        & $7.89\times 10^{-4} $                                                         & $2.10\times 10^{3}$     & 2.54        & 1.40        & $2.95\times 10^{-7}$        & $8.53\times 10^{-4} $                                                         \\
Kepler-186 f  & $9.81\times 10^{6}$     & 1.71        & 1.21        & $7.79\times 10^{-6}$        & $9.12\times 10^{-2}$                                                          & $1.02\times 10^{3}$     & 1.71        & 1.17        & $1.09\times 10^{-5}$        & $1.28\times 10^{-1}$                                                          \\
Kepler-138 d   & $1.29\times 10^{8}$     & $0.64 $       & 1.33        & $9.96\times 10^{-5}$        & $1.73\times 10^{1}$                                                          & $1.09\times 10^{4}$     & $0.64$        & 1.21        & $1.48\times 10^{-4}$        & $2.56\times 10^{1}$                                                          \\
Kepler-1649 c  & $8.28\times 10^{5}$     & 1.20        & 1.12        & $4.44\times 10^{-6}$        & $1.37\times 10^{-1}$                                                          & $3.72\times 10^{4}$     & 1.20        & 1.06        & $4.78\times 10^{-6}$        & $1.47\times 10^{-1}$                                                          \\
Kepler-1652 b  & $2.42\times 10^{5}$     & 3.19        & 1.70        & $1.18\times 10^{-6}$        & $1.85\times 10^{-3}$                                                          & $9.02\times 10^{3}$     & 3.19        & 1.60        & $1.29\times 10^{-6}$        & $2.01\times 10^{-3}$                                                          \\
Kepler-296 e   & $1.90\times 10^{5}$     & 2.96        & 1.61        & $9.28\times 10^{-7}$        & $1.78\times 10^{-3}$                                                          & $6.98\times 10^{3}$     & 2.96        & 1.53        & $1.01\times 10^{-6}$        & $1.93\times 10^{-3}$                                                          \\
Kepler-296 f   & $1.26\times 10^{5}$     & 3.89        & 1.94        & $6.05\times 10^{-7}$        & $5.56\times 10^{-4}$                                                          & $4.57\times 10^{3}$     & 3.89        & 1.80        & $6.59\times 10^{-7}$        & $6.05\times 10^{-4} $                                                         \\
Kepler-705 b   & $2.20\times 10^{5}$     & 5.10        & 2.38        & $9.95\times 10^{-7}$        & $4.42\times 10^{-4}$                                                          & $7.16\times 10^{3}$     & 5.10        & 2.10        & $1.10\times 10^{-6}$        & $4.88\times 10^{-4}$                                                          \\
LHS 1140 b     & $1.08\times 10^{6}$     & 6.98        & 1.81        & $5.43\times 10^{-6}$        & $1.20\times 10^{-3}$                                                          & $4.25\times 10^{4}$     & 6.98        & 1.72        & $5.88\times 10^{-6}$        & $1.30\times 10^{-3}$                                                          \\
Proxima Cen b & $2.96\times 10^{6}$     & 1.27        & 1.14        & $1.60\times 10^{-5}$        & $4.19\times 10^{-1}$                                                          & $1.36\times 10^{5}$     & 1.27        & 1.08        & $1.71\times 10^{-5}$        & $4.50\times 10^{-1}$                                                          \\
Ross 128 b    & $1.71\times 10^{6}$     & 1.40        & 1.17        & $9.15\times 10^{-6}$        & $1.87\times 10^{-1}$                                                          & $7.68\times 10^{4}$     & 1.40        & 1.11        & $9.84\times 10^{-6}$        & $2.01\times 10^{-1} $                                                         \\
TOI-700 d      & $1.19\times 10^{5}$     & 1.57        & 1.19        & $5.91\times 10^{-7}$        & $8.76\times 10^{-3}$                                                          & $4.54\times 10^{3}$     & 1.57        & 1.14        & $6.40\times 10^{-7}$        & $9.48\times 10^{-3}$                                                          \\
TRAPPIST-1 d  & $5.20\times 10^{8}$     & $0.39$        & 1.66        & $1.09\times 10^{-3}$        & $7.04\times 10^{2}$                                                          & $4.41\times 10^{6}$     & $0.39$        & $0.79$        & $2.82\times 10^{-3}$        & $1.82\times 10^{3} $                                                         \\
TRAPPIST-1 e  & $1.55\times 10^{8}$     & $0.69$        & 1.26        & $4.60\times 10^{-4}$        & $6.05\times 10^{1} $                                                         & $2.80\times 10^{6}$     & $0.69$        & $0.92$        & $7.12\times 10^{-4}$        & $9.35\times 10^{1}  $                                                        \\
TRAPPIST-1 f  & $6.97\times 10^{7}$     & 1.04        & 1.20        & $2.24\times 10^{-4}$        & $1.03\times 10^{1}$                                                          & $1.41\times 10^{6}$     & 1.04        & 1.06        & $3.16\times 10^{-4}$        & $1.45\times 10^{1} $                                                         \\
TRAPPIST-1 g  & $4.72\times 10^{7}$     & 1.32        & 1.24        & $1.59\times 10^{-4}$        & $3.77\times 10^{0}$                                                          & $1.05\times 10^{6}$     & 1.32        & 1.14        & $2.17\times 10^{-4}$        & $5.12\times 10^{0}$                                                          \\
Wolf 1061 c    & $2.03\times 10^{6}$     & 3.41        & 1.83        & $9.65\times 10^{-6}$        & $1.26\times 10^{-2}$                                                          & $7.32\times 10^{4}$     & 3.41        & 1.66        & $1.05\times 10^{-5}$        & $1.38\times 10^{-2}$                                                        
\end{tabular}%
}

\caption{Estimated properties of the planets at 0.1 Gyr (M2-M8V), 0.2 Gyr (M0-M1.5V), and at 5 Gyr due to stellar wind stripping. Note that planet mass and radius values at 0.1/0.2 Gyr are not necessarily realistic. They are just representations of the amount of mass accumulated backward in time due to mass lost throughout 5 Gyr. Further percent envelope fraction lost values \textgreater 100\% are also not realistic and indicate that the primordial envelope could be lost several times over.}
\label{tab:p2}
\end{table*}
\end{center}

\begin{center}
\begin{table*}
\resizebox{\textwidth}{!}{%
\begin{tabular}{cllllll|llllll}
\hline
\hline
age           & \multicolumn{6}{c|}{0.1 Gyr}                                                                                                                                                                                                                                                      & \multicolumn{6}{c}{5 Gyr}                                                                                                                                                                                                                                                        \\ \hline
parameter     & \multicolumn{1}{c}{$\dot{m}$}    & \multicolumn{1}{c}{$\rm M_p$}           & \multicolumn{1}{c}{$\rm R_p$}           & \multicolumn{1}{c}{$\rm R_{XUV}$}       & \multicolumn{1}{c}{mass lost}       & \multicolumn{1}{c|}{\begin{tabular}[c]{@{}c@{}}envelope \\ fraction lost\end{tabular}} & \multicolumn{1}{c}{$\dot{m}$}    & \multicolumn{1}{c}{$\rm M_p$}           & \multicolumn{1}{c}{$\rm R_p$}           & \multicolumn{1}{c}{$\rm R_{XUV}$}       & \multicolumn{1}{c}{mass lost}       & \multicolumn{1}{c}{\begin{tabular}[c]{@{}c@{}}envelope \\ fraction lost\end{tabular}} \\ \hline
unit          & \multicolumn{1}{c}{(g $s^{-1}$)} & \multicolumn{1}{c}{($\rm M_{\oplus}$)} & \multicolumn{1}{c}{($\rm R_{\oplus}$)} & \multicolumn{1}{c}{($\rm R_{\oplus}$)} & \multicolumn{1}{c}{($\rm M_{\oplus}$)} & \multicolumn{1}{c|}{($\%$)}                                                            & \multicolumn{1}{c}{(g $s^{-1}$)} & \multicolumn{1}{c}{($\rm M_{\oplus}$)} & \multicolumn{1}{c}{($\rm R_{\oplus}$)} & \multicolumn{1}{c}{($\rm R_{\oplus}$)} & \multicolumn{1}{c}{($\rm M_{\oplus}$)} & \multicolumn{1}{c}{($\%$)}                                                            \\ \hline
GJ 1061 c     & $3.03 \times 10^{12}$                         & 9.98                             & $14.9$                            & $31.3$                            & $9.28\times 10^{0}$                            & $1.05 \times 10^{5}$                                                                               & $3.36 \times 10^{8}$                         & 1.74                             & 1.18                             & 1.75                             & $1.38 \times 10^{1}$                            & $1.56 \times 10^{5}$                                                                              \\
GJ 1061 d     & $9.83 \times 10^{12}$                         & 4.72                             & $14.9$                            & $42.7$                            & $5.20\times 10^{0}$                              & $6.89 \times 10^{4}$                                                                               & $1.21 \times 10^{8}$                         & 1.64                             & 1.16                             & 1.80                             & $9.89\times 10^{0}$                             & $1.31 \times 10^{5}$                                                                              \\
GJ 1132 b      & $5.96 \times 10^{12}$                         & $27.30$                            & $14.6$                            & $20.4$                            & $2.79 \times 10^{1}$                            & $3.36 \times 10^{6}$                                                                               & $2.88 \times 10^{11}$                         & 1.66                            & 1.13                             & 1.70                            & $5.54 \times 10^{1}$                            & $6.68 \times 10^{6}$                                                                              \\
GJ 1214 b      & $1.22 \times 10^{13}$                         & $20.40$                            & $12.20$                            & $22.7$                            & $1.64 \times 10^{1}$                            & $3.68 \times 10^{3}$                                                                               & $5.90 \times 10^{9}$                         & 6.26                             & 2.85                             & 3.29                             & $4.25 \times 10^{1}$                            & $9.52 \times 10^{3}$                                                                              \\
GJ 163 c       & $1.62 \times 10^{9}$                         & 6.81                             & 3.23                             & 4.18                             & $7.35 \times 10^{-3}$                            & $1.55\times 10^{0}$                                                                                 & $5.90 \times 10^{7}$                         & 6.80                             & 2.50                             & 3.11                             & $8.53 \times 10^{-3}$                            & $1.79\times 10^{0}$                                                                               \\
GJ 180 c       & $1.22 \times 10^{9} $                        & 6.41                             & 3.00                             & 3.83                             & $5.89 \times 10^{-3}$                            & $1.44\times 10^{0}$                                                                                 & $4.93 \times 10^{7} $                        & 6.40                             & 2.42                             & 2.95                             & $6.70 \times 10^{-3} $                           & $1.64\times 10^{0}$                                                                               \\
GJ 229 A c    & $1.28 \times 10^{8}$                         & 7.27                             & 2.95                             & 3.50                             & $5.28 \times 10^{-4}$                            & $9.44 \times 10^{-2}$                                                                               & $3.94 \times 10^{6}$                         & 7.27                             & 2.60                             & 2.99                             & $6.25 \times 10^{-4}$                            & $1.12 \times 10^{-1}$                                                                              \\
GJ 273 b      & $1.56 \times 10^{9}$                         & 2.90                             & 2.00                             & 2.77                             & $6.72 \times 10^{-3}$                            & $1.38 \times 10^{1}$                                                                               & $5.13 \times 10^{7}$                         & 2.89                             & 1.51                             & 2.00                             & $7.75 \times 10^{-3}$                            & $1.59 \times 10^{1}$                                                                              \\
GJ 3293 d     & $6.86 \times 10^{8}$                         & 7.60                             & 3.58                             & 4.43                             & $2.87 \times 10^{-3}$                            & $4.63 \times 10^{-1}$                                                                               & $2.17 \times 10^{7}$                         & 7.60                             & 2.67                             & 3.14                             & $3.43 \times 10^{-3}$                            & $5.52 \times 10^{-1}$                                                                              \\
GJ 357 d       & $2.90 \times 10^{8}$                         & 6.10                             & 2.81                             & 3.37                             & $1.55 \times 10^{-3}$                            & $4.31 \times 10^{-1}$                                                                               & $1.38 \times 10^{7}$                         & 6.10                             & 2.34                             & 2.74                             & $1.72 \times 10^{-3}$                            & $4.78 \times 10^{-1}$                                                                              \\
GJ 667 C c    & $5.60 \times 10^{8}$                         & 3.80                             & 2.05                             & 2.59                             & $3.25 \times 10^{-3}$                            & $3.18\times 10^{0}$                                                                                 & $3.00 \times 10^{7}$                         & 3.80                             & 1.78                             & 2.19                             & $3.53 \times 10^{-3} $                           & $3.45\times 10^{0}$                                                                               \\
GJ 667 C e    & $1.10 \times 10^{8}$                         & 2.70                             & 1.59                             & 1.93                             & $7.29 \times 10^{-4}$                            & $1.79\times 10^{0}$                                                                                 & $7.17 \times 10^{6} $                        & 2.70                             & 1.46                             & 1.74                             & $7.76 \times 10^{-4}$                            & $1.90\times 10^{0}$                                                                               \\
GJ 667 C f    & $2.72 \times 10^{8}$                         & 2.70                             & 1.67                             & 2.11                             & $1.61 \times 10^{-3} $                           & $3.95\times 10^{0}$                                                                                 & $1.50 \times 10^{7}$                         & 2.70                             & 1.47                             & 1.80                             & $1.74 \times 10^{-3}$                           & $4.26\times 10^{0}$                                                                               \\
GJ 682 b      & $2.62 \times 10^{8}$                         & 4.40                             & 2.17                             & 2.74                             & $1.62 \times 10^{-3}$                            & $1.07\times 10^{0}$                                                                                 & $1.54 \times 10^{7}$                         & 4.40                             & 1.94                             & 2.39                             & $1.74 \times 10^{-3}$                            & $1.15\times 10^{0}$                                                                               \\
GJ 832 c       & $8.54 \times 10^{8}$                         & 5.40                             & 2.63                             & 3.32                             & $4.63 \times 10^{-3}$                            & $1.77\times 10^{0}$                                                                                 & $4.15 \times 10^{7}$                         & 5.40                             & 2.18                             & 2.71                             & $5.10 \times 10^{-3} $                           & $1.95\times 10^{0}$                                                                               \\
K2-18 b        & $4.72 \times 10^{8}$                         & 8.93                             & 2.83                             & 3.35                             & $2.22 \times 10^{-3}$                            & $2.21 \times 10^{-1}$                                                                               & $2.28 \times 10^{7}$                         & 8.92                             & 2.37                             & 2.73                             & $2.41 \times 10^{-3}$                            & $2.40 \times 10^{-1}$                                                                              \\
K2-288 B b     & $2.79 \times 10^{8}$                         & 4.27                             & 2.13                             & 2.59                             & $1.73 \times 10^{-3}$                            & $1.24\times 10^{0}$                                                                                 & $1.66 \times 10^{7}$                         & 4.27                             & 1.91                             & 2.27                             & $1.87 \times 10^{-3}$                            & $1.34\times 10^{0}$                                                                               \\
K2-3 c         & $2.40 \times 10^{9}$                         & 3.11                             & 2.37                             & 3.67                             & $9.65 \times 10^{-3}$                            & $1.64 \times 10^{1}$                                                                               & $7.12 \times 10^{7}$                         & 3.10                             & 1.77                             & 2.55                             & $1.13 \times 10^{-2}$                            & $1.93 \times 10^{1}$                                                                              \\
K2-3 d         & $3.34 \times 10^{8}$                         & 2.80                             & 1.75                             & 2.32                             & $1.98 \times 10^{-3}$                            & $4.00\times 10^{0}$                                                                               & $1.85 \times 10^{7}$                         & 2.80                             & 1.53                             & 1.95                             & $2.14 \times 10^{-3}$                            & $4.31\times 10^{0}$                                                                               \\
K2-72 e        & $6.48 \times 10^{9}$                         & 2.22                             & 2.30                             & 3.63                             & $1.57 \times 10^{-2}$                            & $6.63 \times 10^{1} $                                                                              & $7.29 \times 10^{7}$                         & 2.21                             & 1.29                             & 1.83                             & $2.78 \times 10^{-2}$                            & $1.17 \times 10^{2}$                                                                              \\
K2-9 b         & $3.94 \times 10^{9}$                         & 5.70                             & 3.00                             & 4.01                             & $1.63 \times 10^{-2}$                            & $5.43\times 10^{0}$                                                                                & $1.23 \times 10^{8}$                         & 5.69                             & 2.26                             & 2.84                             & $1.95 \times 10^{-2}$                            & $6.48\times 10^{0}$                                                                               \\
Kepler-1229 b & $8.31 \times 10^{7}$                         & 2.54                             & 1.52                             & 1.87                             & $5.54 \times 10^{-4}$                            & $1.61\times 10^{0}$                                                                                & $5.49 \times 10^{6}$                         & 2.54                             & 1.40                             & 1.69                             & $5.89 \times 10^{-4}$                            & $1.71\times 10^{0}$                                                                               \\
Kepler-186 f  & $3.22 \times 10^{7}$                         & 1.71                             & 1.26                             & 1.55                             & $2.19 \times 10^{-4}$                            & $2.57\times 10^{0}$                                                                                & $2.19 \times 10^{6}$                         & 1.71                             & 1.17                             & 1.42                             & $2.33 \times 10^{-4}$                            & $2.72\times 10^{0}$                                                                               \\
Kepler-138 d   & $1.50 \times 10^{12}$                         & 8.88                             & $16.5$                            & $41.3$                            & $8.76\times 10^{0}$                              & $1.52 \times 10^{6}$                                                                               & $8.87 \times 10^{10}$                         & 0.64                             & 1.21                             & 1.92                            & $1.14 \times 10^{1}$                            & $1.98 \times 10^{6}$                                                                              \\
Kepler-1649 c  & $2.12 \times 10^{12}$                         & 8.00                             & $15.70$                            & $33.00$                            & $7.52\times 10^{0}$                              & $2.32 \times 10^{5}$                                                                               & $1.86 \times 10^{8}$                         & 1.20                             & 1.06                             & 1.94                             & $1.09 \times 10^{1}$                            & $3.37 \times 10^{5}$                                                                              \\
Kepler-1652 b  & $3.14 \times 10^{8}$                         & 3.19                             & 1.82                             & 2.34                             & $1.88 \times 10^{-3}$                            & $2.95\times 10^{0}$                                                                                & $1.78 \times 10^{7}$                         & $3.19$                             & 1.61                             & 2.01                             & $2.03 \times 10^{-3}$                            & $3.19\times 10^{0}$                                                                               \\
Kepler-296 e   & $4.79 \times 10^{8}$                         & 2.96                             & 1.81                             & 2.50                             & $2.62 \times 10^{-3}$                            & $5.02\times 10^{0}$                                                                                & $2.33 \times 10^{7}$                         & 2.96                             & 1.55                             & 2.04                             & $2.86 \times 10^{-3}$                            & $5.49\times 10^{0}$                                                                               \\
Kepler-296 f   & $1.59 \times 10^{8}$                         & 3.89                             & 1.99                             & 2.50                             & $1.02 \times 10^{-3}$                            & $9.38 \times 10^{-1}$                                                                               & $1.00 \times 10^{7}$                         & 3.89                             & 1.80                             & 2.22                             & $1.09 \times 10^{-3}$                            & $1.00\times 10^{0}$                                                                               \\
Kepler-705 b   & $2.65 \times 10^{8}$                         & 5.10                             & 2.41                             & 2.99                             & $1.56 \times 10^{-3}$                            & $6.92 \times 10^{-1}$                                                                               & $1.46 \times 10^{7}$                         & 5.10                             & 2.10                             & 2.54                             & $1.69 \times 10^{-3}$                            & $7.52 \times 10^{-1}$                                                                              \\
LHS 1140 b     & $2.30 \times 10^{8}$                         & 6.98                             & 1.85                             & 2.21                             & $1.57 \times 10^{-3}$                            & $3.47 \times 10^{-1}$                                                                               & $1.59 \times 10^{7}$                         & 6.38                             & 1.64                             & 2.02                             & $1.67 \times 10^{-3}$                            & $3.68 \times 10^{-1}$                                                                              \\
Proxima Cen b & $2.34 \times 10^{12}$                         & 9.72                             & $15.4$                            & $29.1$                            & $9.27\times 10^{0}$                              & $2.43 \times 10^{5} $                                                                              & $4.76 \times 10^{8}$                         & 1.28                             & 1.08                             & 2.23                             & $1.32 \times 10^{1}$                            & $3.45 \times 10^{5}$                                                                              \\
Ross 128 b    & $2.39 \times 10^{12}$                         & $10.20$                            & $15.3$                            & $32.1$                            & $9.62\times 10^{0}$                              & $1.96 \times 10^{5}$                                                                               & $4.21 \times 10^{8}$                         & 1.40                             & 1.11                             & 2.34                             & $1.39 \times 10^{1}$                            & $2.83 \times 10^{5}$                                                                              \\
TOI-700 d      & $9.27 \times 10^{8}$                         & 1.57                             & 1.66                             & 2.55                             & $3.27 \times 10^{-3}$                            & $4.85 \times 10^{1}$                                                                               & $2.16 \times 10^{7}$                         & 1.57                             & 1.14                             & 1.63                             & $4.06 \times 10^{-3}$                            & $6.01 \times 10^{1}$                                                                              \\
TRAPPIST-1 d  & $2.36 \times 10^{10}$                         & $109.00$                            & $10.40$                            & $11.10$                            & $1.09 \times 10^{2}$                            & $7.02 \times 10^{7}$                                                                               & $2.23 \times 10^{8}$                         & 0.39                            & 0.79                             & 1.10                             & $1.09 \times 10^{2}$                            & $7.02 \times 10^{7}$                                                                              \\
TRAPPIST-1 e  & $2.59 \times 10^{12}$                         & $12.90$                            & $15.30$                            & $26.40$                            & $1.31 \times 10^{1}$                            & $1.73 \times 10^{6}$                                                                               & $1.09 \times 10^{11}$                         & 0.69                             & 0.92                             & 1.39                           & $1.86 \times 10^{1}$                            & $2.45 \times 10^{6}$                                                                              \\
TRAPPIST-1 f  & $2.16 \times 10^{12}$                         & 9.72                             & $15.40$                            & $28.3$                            & $9.44\times 10^{0}$                              & $4.33 \times 10^{5}$                                                                               & $1.44 \times 10^{9}$                         & 1.05                             & 1.05                             & 1.58                             & $1.35 \times 10^{1}$                            & $6.18 \times 10^{5}$                                                                              \\
TRAPPIST-1 g  & $2.57 \times 10^{12}$                         & 7.10                             & $15.4$                            & $31.3$                            & $6.63\times 10^{0}$                              & $1.57 \times 10^{5}$                                                                               & $1.85 \times 10^{8}$                         & 1.32                             & 1.13                             & 1.78                             & $1.04 \times 10^{1} $                           & $2.45 \times 10^{5}$                                                                              \\
Wolf 1061 c    & $2.57 \times 10^{8}$                         & 3.41                             & 1.90                             & 2.51                             & $1.50 \times 10^{-3}$                            & $1.97\times 10^{0}$                                                                                & $1.39 \times 10^{7}$                         & 3.41                             & 1.67                             & 2.13                             & $1.63 \times 10^{-3}$                            & $2.14\times 10^{0}$                                                                              
\end{tabular}%
}

\caption{Estimated properties of the planets at 0.1 Gyr (M2-M8V), 0.2 Gyr (M0-M1.5V), and at 5 Gyr due to photoevaporation stripping. Note here also that planet mass and radius values at 0.1/0.2 Gyr are not necessarily realistic. They are just representations of the amount of mass accumulated backward in time due to mass lost throughout 5 Gyr. Further percent envelope fraction lost values \textgreater 100\% are also not realistic and simply indicate that the primordial envelope could be lost several times over.}
\label{tab:p3}

\end{table*}
\end{center}

\subsection{Close-in orbiting exoplanets}
Based on our estimations, the stellar wind did not drive significant atmosphere loss in the close-in planets Kepler-138d, K2-3c, GJ 1132 b, and GJ 1214 b. Due to uncertainties in the rotation period at younger ages of these stars, we have also computed upper and lower limits to the mass loss for all planets based on the upper and lower errors of the rotation period. 

\begin{figure*}
\begin{center}
 \includegraphics[width=0.48\textwidth]{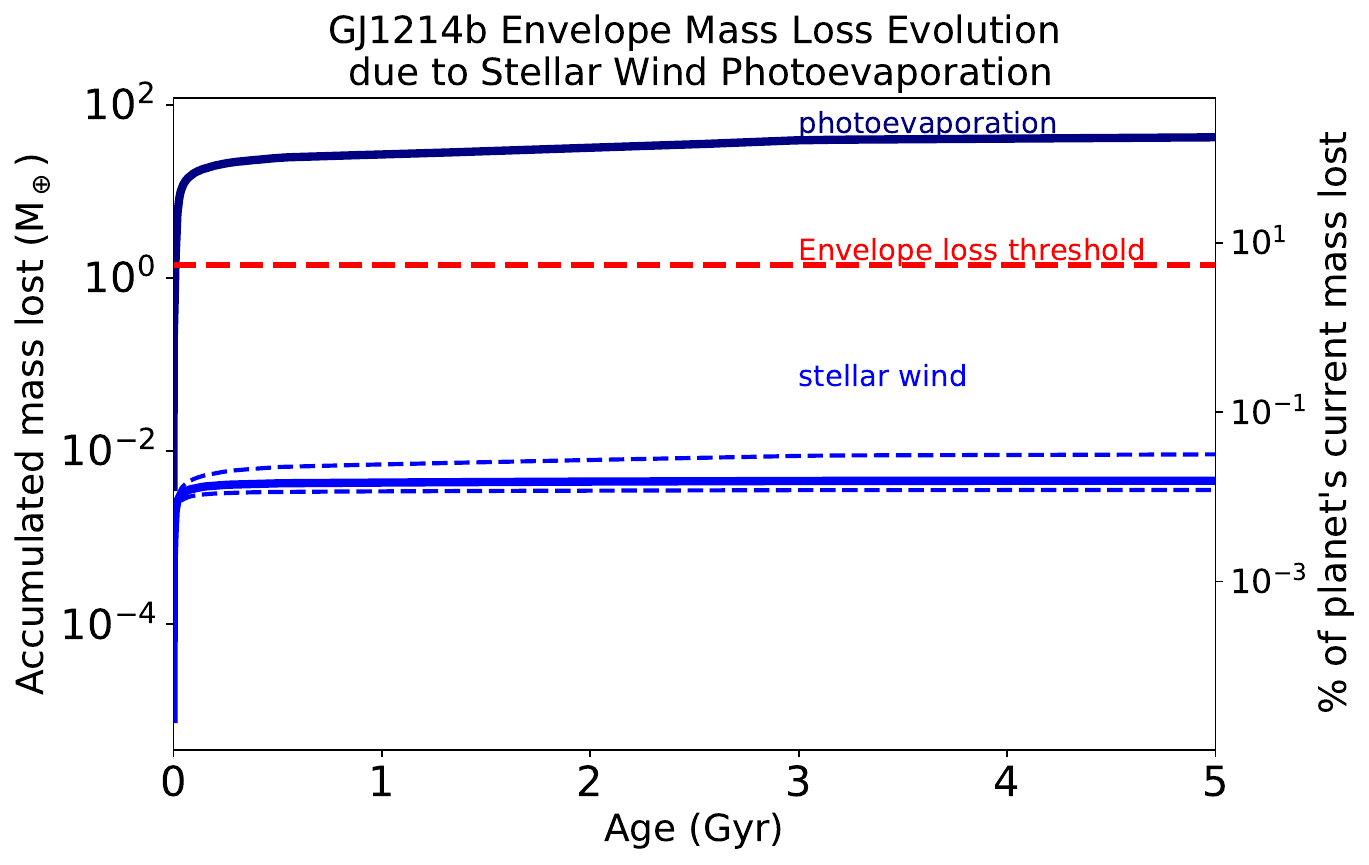}
 \includegraphics[width=0.48\textwidth]{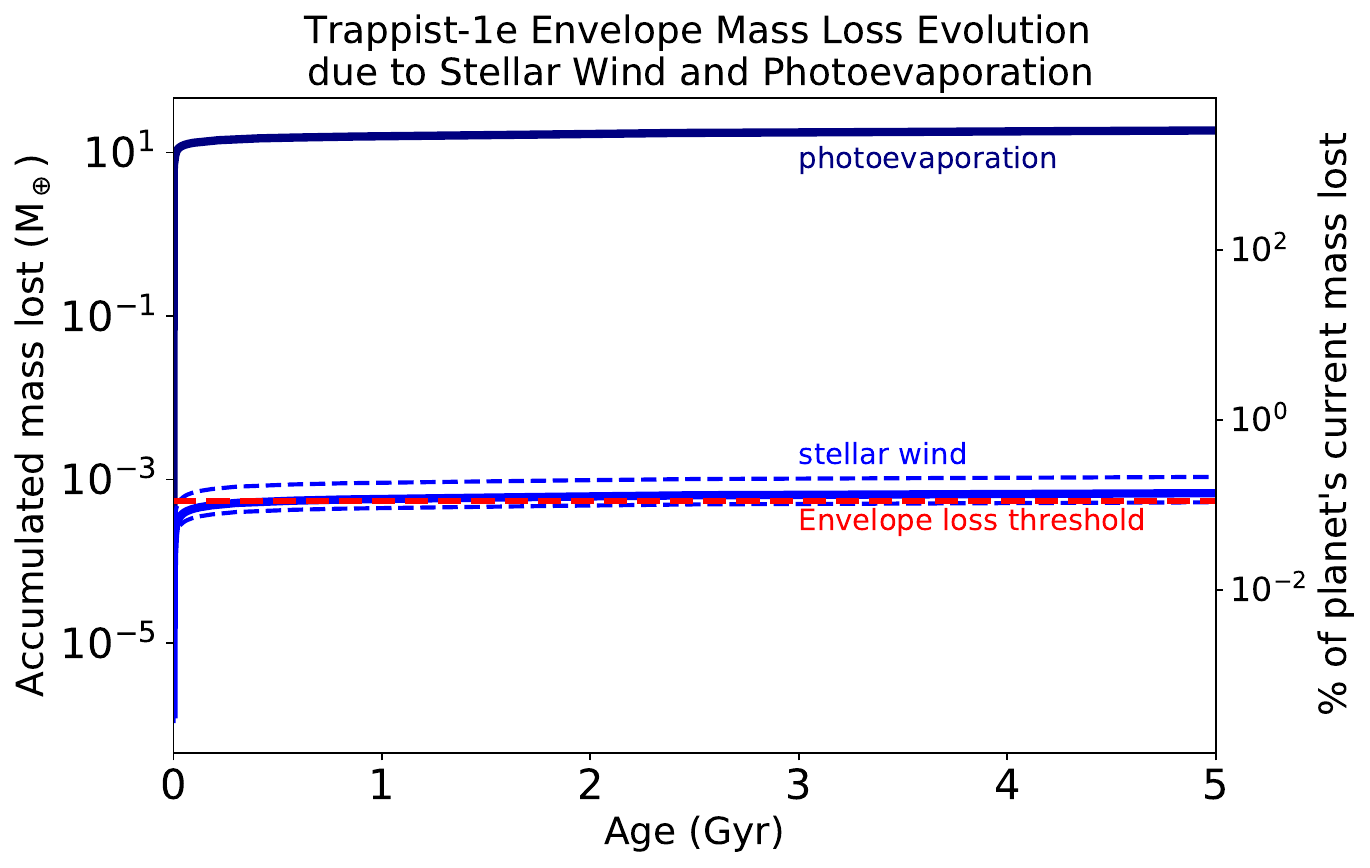}\\
\caption{Sample results for close-in orbiting planet GJ 1214 b and HZ planet Trappist-1e. The accumulated envelope mass loss is shown in blue. The upper and lower limits for stellar wind due to rotation period uncertainties are shown as dashed blue lines. The dashed red line represents the threshold of losing the primordial envelope mass.}
   \label{fig:errors}
\end{center}
\end{figure*}

Even with the upper limit, \textless 10\% of the primordial envelope was lost for all 4 planets due to stellar wind. However,  photoevaporation would drive the loss of 100\% of the atmosphere of GJ 1132 b, GJ 1214 b, and Kepler-138d. A comparison of the mass lost due to stellar wind and photoevaporation for GJ 1214 b is shown in Figure~\ref{fig:errors}. 

\subsection{HZ exoplanets}
\begin{figure}
%\centering
\includegraphics[width=0.5\textwidth]{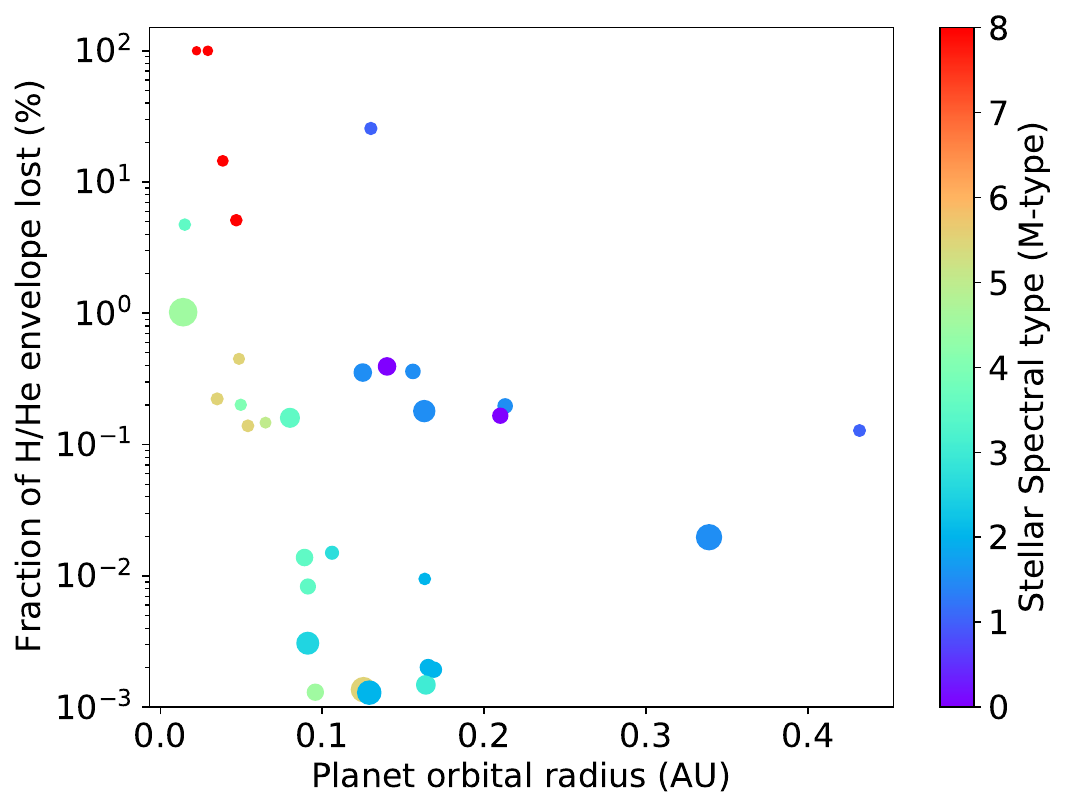}
 \includegraphics[width=0.52\textwidth]{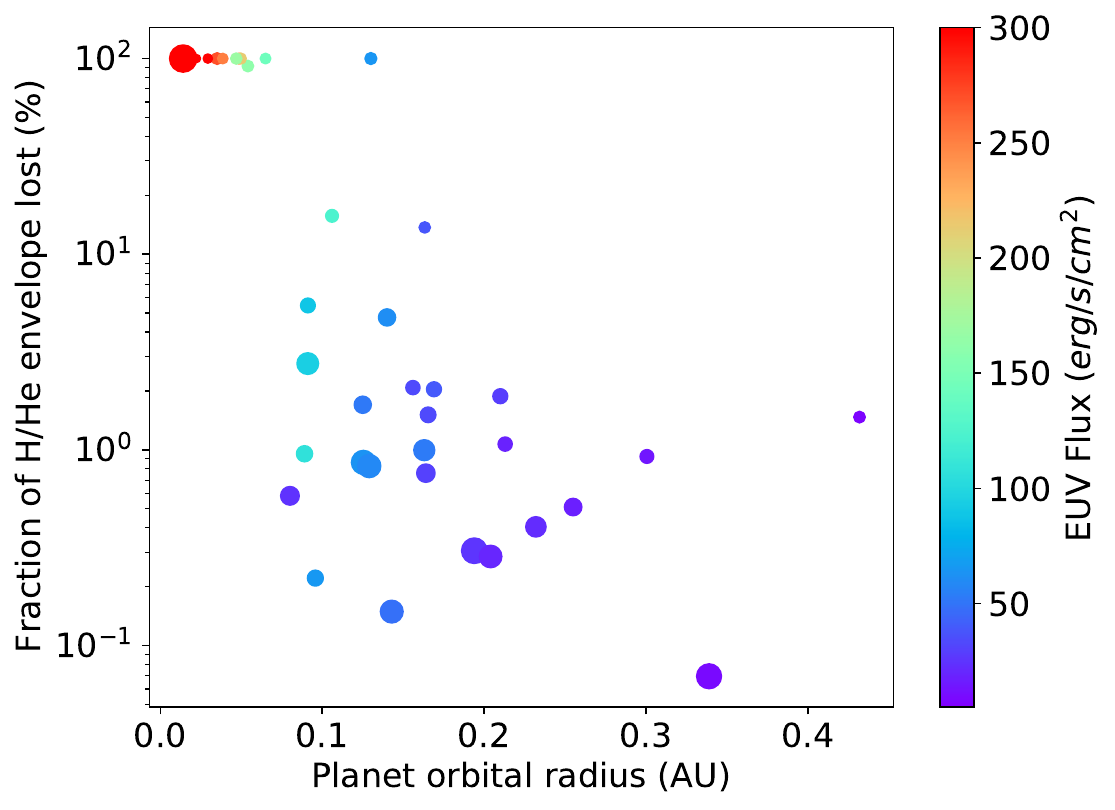}\\
\caption{Envelope fraction lost due to stellar wind (upper) and photoevaporation (lower) as a function of parameters associated with each respective model. EUV flux was calculated at 5 Gyr and at planetary orbit. The size of the scatter points is a function of the radius of the planet.}
   \label{fig:scatter plots}
\end{figure}

\begin{figure}
\centering
\includegraphics[width=0.5\textwidth]{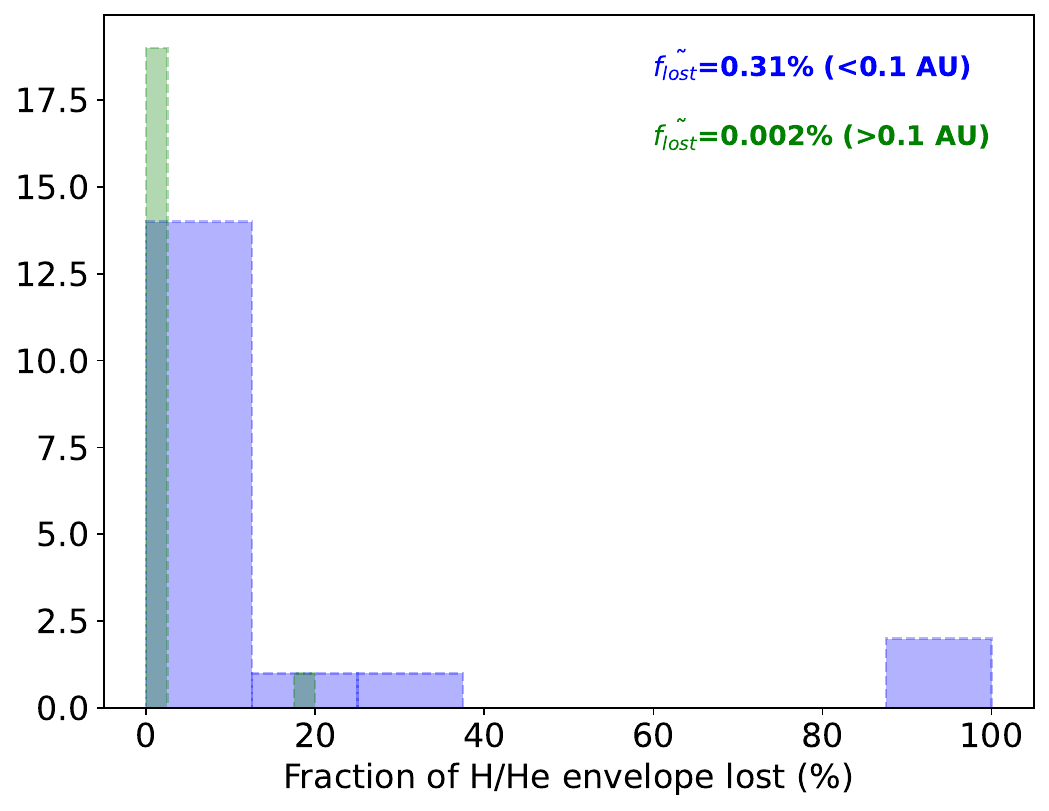}
 \includegraphics[width=0.5\textwidth]{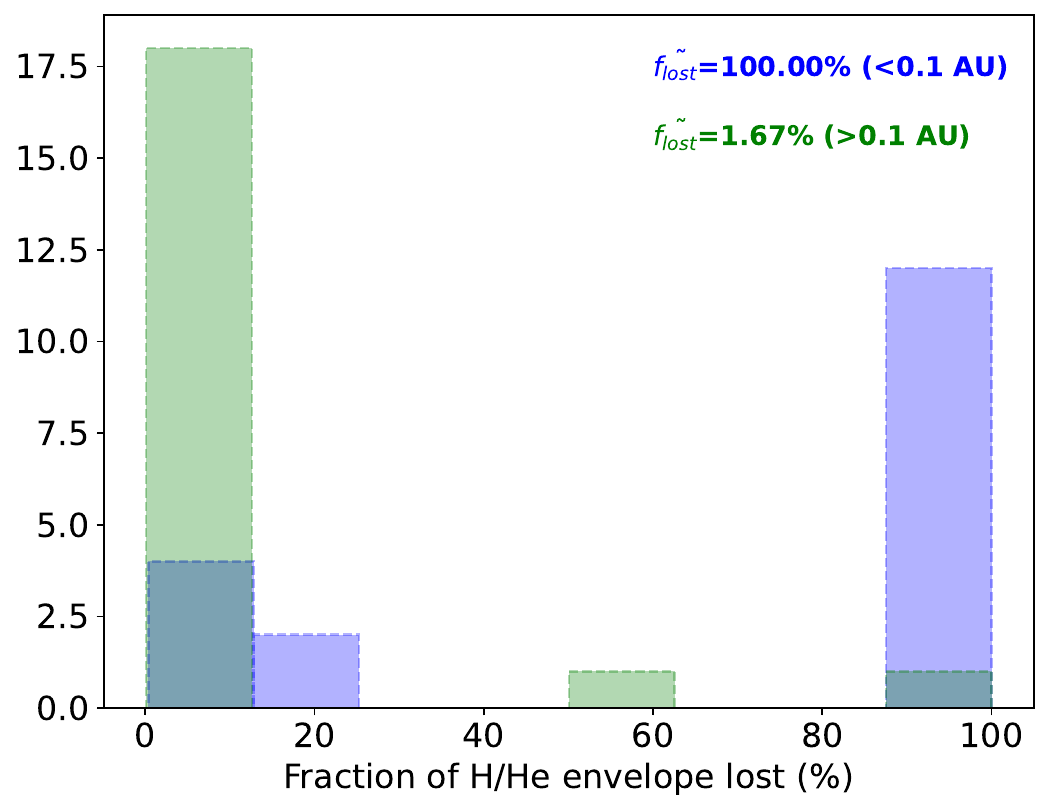}\\
\caption{Histograms of envelope fraction lost due to stellar wind (upper) and photoevaporation (lower). The fraction of envelope lost for planets located below 0.1 AU is indicated in blue and above 0.1 AU, in green, with their respective median values.}
   \label{fig:hist plots}
\end{figure}

For the HZ planets orbiting early-type M dwarfs at $>$ 0.1 AU, photoevaporation was the only mechanism that could significantly strip the estimated primordial envelope. This suggests that planets in the outer HZ of an early-type M-dwarf are unlikely to lose an atmosphere due to stellar wind. For the planets around late-type M dwarfs, orbiting at closer orbit within the HZ, between 0.01-0.03 AU from their host star, both stellar wind and photoevaporation played a non-negligible role in atmospheric stripping, with photoevaporation still being more effective. This is demonstrated in Figure~\ref{fig:scatter plots} where only the planets around M-dwarfs of higher spectral type and in close orbit lost their primordial envelope due to stellar wind. An example of this is the Trappist-1 system, which experienced significant atmospheric stripping due to both photoevaporation and stellar wind due to the close orbit of the planets. Overall, photoevaporation played a significant role in atmospheric escape for HZ planets. Particularly those in the inner HZ, as shown in Fig.~\ref{fig:hist plots}, where the median fraction of envelope loss for planets orbiting \textless 0.1 AU was 100\%. The results for the atmospheric mass loss due to photoevaporation and stellar wind at 5 Gyr in the planet's lifetime are shown in Table~\ref{tab:comp}.

\begin{table}
\centering
\begin{tabular}{ccc}
\hline
\hline
planet        & \begin{tabular}[c]{@{}c@{}}Envelope lost due \\ to photoevaporation (\%)\end{tabular} & \begin{tabular}[c]{@{}c@{}}Envelope lost due \\ to stellar wind (\%)\end{tabular} \\ \hline
GJ 1061 c     & 100.00                                                                                & 0.22                                                                              \\
GJ 1061 d     & 100.00                                                                                 & 0.14                                                                              \\
GJ 1132 b      & 100.00                                                                                & 4.74                                                                              \\
GJ 1214 b      & 100.00                                                                                & 1.02                                                                              \\
GJ 163 c       & 1.79                                                                                  & 0.00                                                                              \\
GJ 180 c       & 1.64                                                                                  & 0.00                                                                              \\
GJ 229 A c    & 0.11                                                                                  & 0.02                                                                              \\
GJ 273 b      & 15.90                                                                                  & 0.01                                                                              \\
GJ 3293 d     & 0.55                                                                                  & 0.00                                                                              \\
GJ 357 d       & 0.48                                                                                  & 0.00                                                                              \\
GJ 667 C c    & 3.45                                                                                  & 0.35                                                                              \\
GJ 667 C e    & 1.90                                                                                  & 0.20                                                                              \\
GJ 667 C f    & 4.26                                                                                  & 0.36                                                                              \\
GJ 682 b      & 1.15                                                                                  & 0.16                                                                              \\
GJ 832 c       & 1.95                                                                                  & 0.18                                                                              \\
K2-18 b        & 0.24                                                                                  & 0.00                                                                              \\
K2-288 B b     & 1.34                                                                                  & 0.00                                                                              \\
K2-3 c         & 19.3                                                                                  & 0.39                                                                              \\
K2-3 d         & 4.31                                                                                  & 0.17                                                                              \\
K2-72 e        & 100.00                                                                                 & 0.01                                                                              \\
K2-9 b         & 6.48                                                                                  & 0.00                                                                              \\
Kepler-1229 b & 1.71                                                                                  & 0.00                                                                              \\
Kepler-186 f  & 2.72                                                                                  & 0.13                                                                              \\
Kepler-138 d   & 100.00                                                                                & 25.64                                                                             \\
Kepler-1649 c  & 100.00                                                                                & 0.15                                                                              \\
Kepler-1652 b  & 3.19                                                                                  & 0.00                                                                              \\
Kepler-296 e   & 5.49                                                                                  & 0.00                                                                              \\
Kepler-296 f   & 1.00                                                                                  & 0.00                                                                              \\
Kepler-705 b   & 0.75                                                                                  & 0.00                                                                              \\
LHS 1140 b     & 0.37                                                                                  & 0.00                                                                              \\
Proxima Cen b & 100.00                                                                                & 0.45                                                                              \\
Ross 128 b    & 100.00                                                                                & 0.20                                                                              \\
TOI-700 d      & 60.11                                                                                 & 0.01                                                                              \\
TRAPPIST-1 d  & 100.00                                                                                & 100.00                                                                            \\
TRAPPIST-1 e  & 100.00                                                                                & 100.00                                                                            \\
TRAPPIST-1 f  & 100.00                                                                                & 14.47                                                                             \\
TRAPPIST-1 g  & 100.00                                                                                & 5.12                                                                              \\
Wolf 1061 c    & 2.14                                                                                  & 0.01                                                                              \\ \hline
\end{tabular}
\caption{Comparison of stellar wind and photoevaporation at 5 Gyr}
\label{tab:comp}
\end{table}

%Photoevaporation is also more effective on planets in close in orbits and that receive high EUV radiation as shown in Figure 10. 

\section{Discussion}

% Even if the primary accreted atmosphere or first-generation outgassed secondary atmosphere is quickly lost during the M dwarf youth, planets may outgas secondary atmospheres after the loss of a primordial atmosphere. The balance between mass loss, accretion, and outgassing will control the composition and mass of each planet atmosphere.

The main result of the preceding analysis is that considering the maximum entrainment of 0.3, stellar wind alone of M-type stars that experience significant decay in time is not sufficient to strip the atmospheres of HZ planets completely. Therefore, processes such as photoevaporation could play a key role in leading to a significant amount of H/He escaping in the first Myrs of the planet's lifetime. Planets at $<$0.1 AU, generally in the inner HZ region, and with H/He envelope fractions of $\sim$0.1\% of the planet's total mass are more susceptible to complete atmospheric loss. We reach similar findings as \cite{owen2016habitability}, which report that at the outer edge of the HZ, evaporation cannot remove a H/He envelope, assuming all planets started with an envelope fraction of 1 percent. Although this paper only focuses on escape due to stellar wind or photoevaporation, many studies have noted that other factors, such as CMEs, can induce ion pickup and enhance atmospheric erosion \citep{lammer2007coronal, cherenkov2017influence}. It is expected that CME rates in M-dwarfs are high \citep{gudel2007sun}. 

Another important factor not incorporated into this study that can enhance atmospheric escape is core-powered mass loss which has been shown to explain the bimodal radius distribution of small close-in exoplanets \citep{ginzburg2018core, gupta2019sculpting, gupta2020signatures}. Outflow due to core-powered mass loss is especially significant for low-mass planets which accrete lighter envelopes ($M_{atm}/M_{core} < 5\%$). For these planets, which comprise a large percent of our sample, core-powered mass loss can dominate over photoevaporation at the early stages of evolution.

It has also been suggested that the presence of a magnetosphere can shield the planet from the charged particles of the stellar wind and prevent erosion \citep{see2014effects}. Here we have not considered the role of the presence of a planet's magnetosphere in the atmospheric escape. The planets studied in this research orbit closely to their host star, indicating that they are more likely to have a weak or non-existent magnetic field due to tidal locking \citep{griebetameier2005cosmic}. Also, observations of  Solar System planets show that the mass escape rate of Earth is similar to the unmagnetized planets Venus and Mars \citep{gunell2018intrinsic}, bringing into question the role of magnetospheres in atmospheric escape. Further, as shown by \cite{gunell2018intrinsic}, ion escape rates may be higher for magnetized planets through the polar caps and cusps. Therefore, more investigation is needed to determine the magnetosphere's influence on atmospheric escape, and should be addressed in future works.

In the following subsections, we detail the implications of our results for the 8 planets that have observations of their transmission spectra taken with HST: GJ-1132b, LHS 1140b, K2-18b, GJ 1214 b, and Trappist-1d-g.

\subsection{GJ-1132b}
For this planet, we estimated a H/He mass loss rate due to stellar wind of $\approx$ $7.65\times 10^{7}$ g s$^{-1}$ for ages of 0.1 Gyr. The mass loss rates from the stellar wind suggest a scenario in which the initial  H/He envelope of GJ 1132 b was not completely stripped over the planet's lifetime. However, the mass loss rate due to photoevaporation of $5.96\times 10^{12}$ g s$^{-1}$ at 0.1 Gyr implies that the planet will be stripped of its primordial envelope in ~100 Myr. Observational work by \cite{swain2021detection} found spectral signatures of aerosol scattering, HCN, and CH$_4$ which suggests that this planet re-established a secondary atmosphere that could be due to planetary outgassing. Our results for photoevaporation are compatible with \cite{estrela2020evolutionary}, which also indicates that GJ 1132 b would have its envelope completely stripped within 100 Myrs.

\subsection{LHS 1140b}
At 0.1 Gyr, the mass loss rate of LHS 1140b due to stellar wind and photoevaporation are $\approx$ $1.08\times 10^{6}$ and $2.30\times 10^{8}$ g s$^{-1}$, respectively, resulting in low escape due to both mechanisms. Previous studies calculated large surface gravities on the planet, indicating a greater chance for atmospheric retention \citep{luger2015extreme}. Current observations indicate the presence of water vapor in its atmosphere \citep{edwards2020hubble}. Because a significant amount of mass was not lost, it is suggestive that pressures on the planet's surface are incompatible with liquid water to exist, only vapor. 

\subsection{K2-18b}
The calculated mass loss rate due to stellar wind is $\approx$ $8.76\times 10^{5}$ g s$^{-1}$ for this planet at 0.1 Gyr. Thus, this low mass loss rate at early ages was not sufficient to strip the calculated primordial envelope, indicating that K2-18b  is currently shrouded by a H/He envelope. \cite{dos2020high} reported a mass loss rate of $\approx$ $1\times 10^{8}$ g s$^{-1}$ due to EUV photoevaporation, comparable to our value of $4.72\times 10^{8}$ g s$^{-1}$ at 5 Gyr, also suggesting that the planet would retain its volatile-rich atmosphere.  Further results from retrieval modeling from other authors show that the data is best matched with an H-dominated atmosphere today with traces of H$_{2}$O vapor \citep{benneke2019water}.

%our results are also consistent with previous studies suggesting a primordial envelope \citep{madhusudhan2020interior,scheucher2020consistently}.

\subsection{GJ 1214 b}
The mass loss rate of this planet due to stellar wind computed in this study at 0.1 Gyr is high at $\approx 1.2\times 10^{9}$ g s$^{-1}$, with values at 5 Gyr lowering to $\approx$ $1.7\times 10^{7}$ g s$^{-1}$. These values are lower by 4 orders of magnitudes than that of photoevaporation with a mass loss rate of $\approx1.22 \times 10^{13}$ g s$^ {-1}$ at 0.1 Gyr, significantly lowering to $\approx 5.90\times 10^{9}$ g s$^{-1}$ at 5 Gyr. %Further, the mass loss rate at 5 Gyr is similar to the hydrodynamic escape rate of 9\times 10^{8 g/s \cite{charbonneau2009super} predicted.
Due to the large initial envelope fraction ($\approx5.5\%$), GJ 1214 b did not lose a H/He envelope due to stellar wind but was predicted to have its atmosphere completely stripped in the first 0.1 Gyr by photoevaporation. \cite{kasper2020nondetection} reported a non-detection of He absorption in the atmosphere of GJ 1214; however, the recent study of \cite{orell2022tentative} reported a tentative detection of He I. 

\subsection{Trappist-1 system (planets d, e, f, and g)}
With the stellar wind and photoevaporation models, planets d and e lost 100\% of their primordial envelope. Planets f and g lost 14\% and 5\% of their primordial envelope due to stellar wind, respectively, and 100\% due to photoevaporation. These results are supported by other photoevaporative models, which predict that all planets would lose their primordial envelope due to photoevaporation \citep{hori2020trappist, turbet2020review}. Current spectroscopic observations rule out cloud-free hydrogen-dominated atmospheres for TRAPPIST-1 d, e, and f, with the significance of 8$\sigma$, 6$\sigma$, and 4$\sigma$, respectively. However, an H-dominated atmosphere could not be ruled out in planet g \citep{de2018atmospheric}.  

\section{Conclusion} 
The wind velocity and density for thirty-eight M-dwarf host stars were calculated with the evolution of the stellar properties taken into account. Rotation period and X-ray flux scaling over time were utilized to estimate the evolution of stellar activity properly and, therefore, of the stellar wind. The X-ray flux was also scaled to determine the mass loss rate due to photoevaporation. These values were computed over a timescale of 5 Gyr. The mass loss rate was computed as a function of time and then accumulated until 5 Gyr to determine the total atmospheric mass loss of the planet's primordial, H/He-dominated atmosphere. Like photoevaporation, our model indicates that stripping due to stellar winds is most effective over the first 0.1 Gyr when the host star is expected to be most active.

We find that close-in orbiting planets, GJ 1132 b, GJ 1214 b, and Kepler-138 d, and the planets within the HZ, GJ 1061 c, d, K2-72 e, Kepler-1649 c, Proxima Cen b, Ross 128 b, and TRAPPIST-1 d-g could completely lose the estimated primordial envelope.  Further, our model results, in conjunction with HST data, indicate that the observed current atmospheres on TRAPPIST-1d-f and GJ 1132 b could be a wispy secondary atmosphere. The lower density of the planet GJ 1214 ($\rm \rho_p = 2.20 \pm 0.16$ $\rm g cm^{-3}$) suggests that the planet could still have a hydrogen envelope that could be explained by outgassing from rocky material \citep{rogers2010three} if, as according to our photoevaporation model, the planet did not manage to retain its primordial envelope.

Our results show that planets orbiting earlier type M-dwarf stars at a distance of \textgreater 0.1 AU are more likely to preserve their atmosphere with median envelope fractions lost of 0.0\% and 1.7\% for stellar wind and photoevaporation, respectively. The rotation period decay for later type M-dwarfs (\textgreater M6) has not been determined; however, it is likely that they do not follow the same trends as earlier spectral type stars and remain highly active for longer periods of time, making orbiting planets more susceptible to mass loss due to stellar wind as seen with our reported results for the Trappist-1 planets. 

More observations are needed to better constrain the evolution of M-dwarf stars' activity (e.g., age-rotation period relationship) and EUV emissions. These properties of the star are crucial for improving the accuracy in the estimation of planetary atmospheric mass loss. Space missions such as the James Webb Space Telescope will improve our understanding of the atmospheres of these planets. The detection of any spectral features in small planets, like the ones in our sample, would be evidence of a low molecular weight atmosphere. These observations will constrain our modeling predictions and pave the path toward searching for life on cooler, habitable zone exoplanets. 

\section*{Acknowledgments} 
This research made use of data published on the NASA Exoplanet Archive operated by the California Institute of Technology, under contract with the National Aeronautics and Space Administration under the Exoplanet Exploration Program. Part of the research was carried out at the Jet Propulsion Laboratory, California Institute of Technology, under a contract with the National Aeronautics and Space Administration (80NM0018D0004).

\section*{Data Availability}
The data supporting this study's findings are available from the corresponding author, Ashini Modi, upon reasonable request.



\bibliographystyle{mnras}
\bibliography{mnras_template} % if your bibtex file is called example.bib

\bsp	% typesetting comment
\label{lastpage}
\end{document}